\documentclass{aa}
\usepackage{graphicx}
\usepackage{txfonts}
\usepackage{natbib}
\usepackage{aalongtable}
\usepackage{url}
\bibpunct{(}{)}{;}{a}{}{,}

\begin{document}

\title{GaBoDS: The Garching-Bonn Deep Survey } \subtitle{V. Data
  release of the ESO Deep-Public-Survey\thanks{Based on observations
    made with the European Southern Observatory telescopes obtained
    from the ESO/ST-ECF Science Archive Facility.}}
\titlerunning{GaBoDS V. - ESO Deep-Public-Survey}

   \author{H. Hildebrandt
          \inst{1}
          \and
          T. Erben\inst{1}
          \and
          J.\,P. Dietrich\inst{1}
          \and
          O. Cordes\inst{1}
          \and
          L. Haberzettl\inst{2}
          \and
          M. Hetterscheidt\inst{1}
          \and
          M. Schirmer\inst{3}
          \and
          O. Schmithuesen\inst{2}
          \and
          P. Schneider\inst{1}
          \and
          P. Simon\inst{1}
          \and
          C. Trachternach\inst{2}
          }

   \offprints{H. Hildebrandt \email{hendrik@astro.uni-bonn.de}}

   \institute{Argelander-Institut f\"ur Astronomie, Universit\"at Bonn\thanks{Founded by merging of the Sternwarte, Radioastronomisches Institut and Institut f\"ur Astrophysik und Extraterrestrische Forschung der Universit\"at Bonn}, Auf dem H\"ugel 71, D-53121 Bonn, Germany
         \and
             Astronomisches Institut der Ruhr-Universit\"at-Bochum,
             Universit\"atsstr. 150, D-44780 Bochum, Germany
         \and
             Isaac Newton Group of Telescopes, Apartado de correos 321, 38700 Santa Cruz de La Palma, Tenerife, Spain\\
             }
   \date{Received ; accepted }
   
   \abstract 
   {}
   {In this paper the optical data of the ESO Deep-Public-Survey
     observed with the Wide Field Imager and reduced with the THELI
     pipeline are described.}  
   {Here we present 63 fully reduced and stacked images. The
     astrometric and photometric calibrations are discussed and the
     properties of the images are compared to images released by the
     ESO Imaging Survey team covering a subset of our data.}
   {These images are publicly released to the community.  Our main
     scientific goals with this survey are to study the high-redshift
     universe by optically pre-selecting high-redshift objects from
     imaging data and to use VLT instruments for follow-up
     spectroscopy as well as weak lensing applications.}
   {}


   \maketitle

\section{Introduction}
\label{sec:introduction}
The ESO Deep-Public-Survey (DPS) is a multi-colour imaging survey
carried out by the ESO Imaging Survey (EIS) team under the program IDs
164.O-0561 and 169.A-0725. It consists of optical data in the
$UBVRI$-bands observed with the Wide Field Imager (WFI) at the 2.2\,m
telescope at La Silla and infrared data in the $J$- and $Ks$-bands
observed with SOFI at the New Technology Telescope. In this paper we
present 63 reduced, photometrically and astrometrically fully
calibrated, and stacked images of the optical part which are released
to the scientific community.\footnote{The images can be downloaded at
  \url{http://marvin.astro.uni-bonn.de/DPS/} and are available via the
  ESO archive
  \url{http://archive.eso.org/archive/eso_data_products.html}.} These
images were reduced and calibrated with our THELI reduction pipeline
\citep{2005astro.ph..1144E}.

The main scientific driver for the ESO DPS were searches for
high-redshift galaxies, distant clusters, high-redshift QSOs, low
surface-brightness galaxies, and gravitational lensing studies. The
survey was designed in a way to deliver a unique dataset also for
studies on Galactic structure, very low-metallicity stars, white
dwarfs, M-dwarfs, and field brown dwarfs.  The deep imaging data can
be used to pre-select objects by colour for follow-up spectroscopy
with VLT instruments.

It was intended to cover an area of three square degrees in three well
separated regions of one square degree each at high galactic latitude
called Deep1, Deep2, and Deep3. Each region consists of four adjacent
WFI pointings ($34'\times33'$) at the same declination named a, b, c,
d in order of decreasing right ascension. Table~\ref{tab:pos}
summarises the positions of the twelve DPS fields. The Deep1 region
was chosen to overlap with the ATESP radio survey
\citep{2000A&AS..146...31P}. The Chandra Deep Field South (CDFS) is
included in the Deep2 region (centred on the field Deep2c), and Deep3
is a random, empty, high galactic latitude field positioned in such a
way that DPS observations are possible over the whole calendar year.

\begin{table}
  \caption{Positions and available colours of the twelve DPS fields and the two mispointings (Deep1e and Deep1f; see text). The region Deep1 overlaps with the ATESP radio survey \citep{2000A&AS..146...31P} and the field Deep2c is centred on the Chandra Deep Field South.}
\label{tab:pos}
\begin{tabular}{ r c c c }
\hline
\hline
field & RA [h m s] & DEC [d m s]& avail.  \\
      & J2000.0    & J2000.0    & colours \\
\hline
Deep1a     & 22:55:00.0 & $-$40:13:00 & $UBVRI$   \\
     b     & 22:52:07.1 & $-$40:13:00 & $UBVRI$  \\
     c     & 22:49:14.3 & $-$40:13:00 & $VRI$     \\
     d     & 22:46:21.4 & $-$40:13:00 & ---      \\
Deep2a     & 03:37:27.5 & $-$27:48:46 & $R$      \\
     b     & 03:34:58.2 & $-$27:48:46 & $UBVRI$  \\
     c     & 03:32:29.0 & $-$27:48:46 & $UBVRI$  \\
     d     & 03:29:59.8 & $-$27:48:46 & $R$      \\
Deep3a     & 11:24:50.0 & $-$21:42:00 & $UBVRI$  \\
     b     & 11:22:27.9 & $-$21:42:00 & $UBVRI$  \\
     c     & 11:20:05.9 & $-$21:42:00 & $UBVRI$  \\
     d     & 11:17:43.8 & $-$21:42:00 & $BVRI$   \\
\hline
Deep1e     & 22:47:47.9 & $-$39:31:06 & $URI$    \\
     f     & 22:44:58.4 & $-$39:31:54 & $I$      \\
\hline
\end{tabular}
\end{table}

Our group has developed a wide-field imaging reduction pipeline and
decided to reduce the DPS data in late 2003 as an ideal test case of a
unique, large dataset fitting our scientific goals. We are mainly
interested in searches for Lyman-break galaxies (LBGs) and
weak-lensing studies supported by photometric redshifts. In November
2004 the EIS team released 40 reduced images of the optical part of
the DPS which enabled us to compare the performance of our pipeline to
a different one. The released EIS data cover a subset of our
data-release and are available via the ESO archive.

The paper is organised as follows. In Sect.~\ref{sec:raw-data} the
instrument, its photometric system, the observing strategy, and the
raw data are described. Sect.~\ref{sec:data_reduction} gives a short
summary of the data reduction with the THELI pipeline with emphasis on
steps which are important for the end user. The properties of our
released images are presented in Sect.~\ref{sec:released-data}, and in
Sect.~\ref{sec:EIS_compare} these properties are compared to the EIS
data release of the ESO DPS. A summary of the data release and an
outlook on projects carried out with the DPS data is given in
Sect.~\ref{sec:summary}.

\section{The raw data}
\label{sec:raw-data}

All optical data of the DPS were observed with the Wide Field Imager
(WFI). This multi-chip, focal-reducer CCD camera has a field of view
of $34'\times33'$ with a filling factor of $\sim96\%$. Seven
broad-band filters are used for the DPS some of which are very
different from the standard Johnson-Cousins $UBVRI$ filters. Their
properties are summarised in Table~\ref{tab:WFI_filters}. The $U$-band
filter ESO \# 841 and the $I$-band filter ESO \# 845 were replaced by
new filters for DPS observations after some time. In
Table~\ref{tab:exp_time} the scheduled exposure times and expected
limiting magnitudes are shown which were chosen to suffice the
scientific goals as described above. For our reduction of the DPS we
used all the raw data available until December 2005 including images
from the two large programmes 164.O-0561 and 169.A-0725 (P.I.
Krautter for both) and, moreover, data from the programmes 67.A-0244
(P.I.  Schneider) and 75.A-0280 (P.I.  Hildebrandt). For the field
Deep2c we added also data from the programmes 168.A-0485 (P.I.
Cesarsky), 64.H-0390 (P.I.  Patat), 66.A-0413 (P.I.  Clocchiatti),
68.D-0273 (P.I.  Cappellaro), 68.A-0443 (P.I.  Clocchiatti), 70.A-0384
(P.I.  Vanzi), and 74.A-9001 (P.I. Kuijken), and we used the WFI
commissioning data and data from the COMBO17 survey (P.I.
Meisenheimer) described in \cite{2004A&A...421..913W}.

\begin{table}
  \caption{Characteristics of the filters used for the DPS. For the computation
    of the effective wavelength and the filter's width the CCD efficiency is included.}
\label{tab:WFI_filters}
\begin{tabular}{ l c c r r }
  \hline
  \hline
  Filter    &ESO \# & eff. WL [$\AA$] & width [$\AA$] & AB corr. [mag]\\
  \hline
  $U\_35060$& 877   & 3429        & 705 & 1.1\\
  $U\_38$   & 841   & 3647        & 373 & 0.8\\
  $B$       & 842   & 4589        & 871 & $-0.1$\\
  $V$       & 843   & 5377        & 793 & 0.0\\
  $R$       & 844   & 6504        & 1502& 0.2\\
  $I$       & 845   & 8635        & 1425& 0.5\\
  $I\_EIS$  & 879   & 8057        & 1402& 0.5\\
\hline
\end{tabular}
\end{table}

\begin{table}
  \caption{Scheduled exposure times and expected limiting magnitudes for the images in
    the different filters. The limiting magnitudes correspond to $5\sigma$ sky
    level in an aperture of $2\arcsec$.}
\label{tab:exp_time}
\begin{tabular}{ l r c }
\hline
\hline
Filter & exp. time [s] & mag lim. [Vega mag]\\
\hline
$U\_35060$   &  $43\,200$ & 25.7\\
$U\_38$    &  $61\,200$ & 26.0\\
$B$    &  $12\,600$ & 26.1\\
$V$    &  $ 9000$ & 26.0\\
$R$    &  $ 9000$ & 26.1\\
$I$    &  $27\,000$ & 25.5\\
$I\_EIS$& $27\,000$ & 25.5\\
\hline
\end{tabular}
\end{table}

Unfortunately, the optical observations of the DPS were not finished.
Not all fields were observed to the scheduled depths in all filters.
Furthermore, some observations were executed under bad sky conditions
so that these images are excluded from the data reduction. There are
also some mispointings present in the DPS raw data for which the FITS
header contains the correct coordinates of one DPS field while the
actually observed area is offset by some arcminutes. Some fields
(especially the field Deep2c centred on the Chandra Deep Field South)
were observed by different programmes to much greater depth, and we
also included these data in our reduction. The estimates of the
limiting magnitudes in Table~\ref{tab:exp_time} were quite
optimistically calculated at a time when the WFI did not yet exist.
The final images are significantly shallower (see
Table~\ref{tab:DPS_images}). For all these reasons the DPS is rather
heterogeneously deep, and this fact should be kept in mind when
dealing with data from different DPS fields.

The desired depth usually requires observations in different nights
for every field in each colour. For many fields, observations spread
over more than a year and images taken under very different
photometric conditions have to be combined.  Even the instrument setup
(baffling, camera rotation etc.) changed over the time of DPS
observations. Furthermore, to reach a final coadded image with an
exposure time as uniform as possible over the field and in order to
get good relative astrometry and photometry between the chips, a wide
dither pattern was chosen for the DPS. The whole observing strategy
(dither pattern, exposure times of the individual images etc.) was
adapted and optimised over the years of DPS observations.

\section{Data reduction}
\label{sec:data_reduction}
The data reduction was performed with our THELI pipeline described in
detail in \citet{2005astro.ph..1144E}. All the raw data were requested
from the ESO Science Archive Facility.

\subsection{Pre-reduction}
The data were sorted in so-called observation runs which include
science, calibration, and standard star images from some adjacent
nights. In the $UBVR$-bands an observation run can easily contain data
from one ($R$) or even two weeks ($UBV$).  The strong fringe patterns
in the $I$-band are variable from night to night and even within one
night, so that $I$-band runs usually contain data from only one or two
hours of observation.

On this run basis, the science images were pre-reduced which includes
overscan correction, bias subtraction, flat-fielding, super-flatting,
and in the $R$- and $I$-band also fringe-removal. No correction for
the inhomogeneous illumination \citep[see][]{2004AN....325..299K} was
applied.  Weight images were created containing flags for bad pixels,
bad columns, and other image defects like satellite tracks which were
masked out by hand. The Landolt standard star images which are
available for most nights were reduced in the same way as the science
images.

\subsection{Absolute photometric calibration}
\label{sec:abs_phot}
\begin{figure*}[t!]
\resizebox{\hsize}{!}{\includegraphics{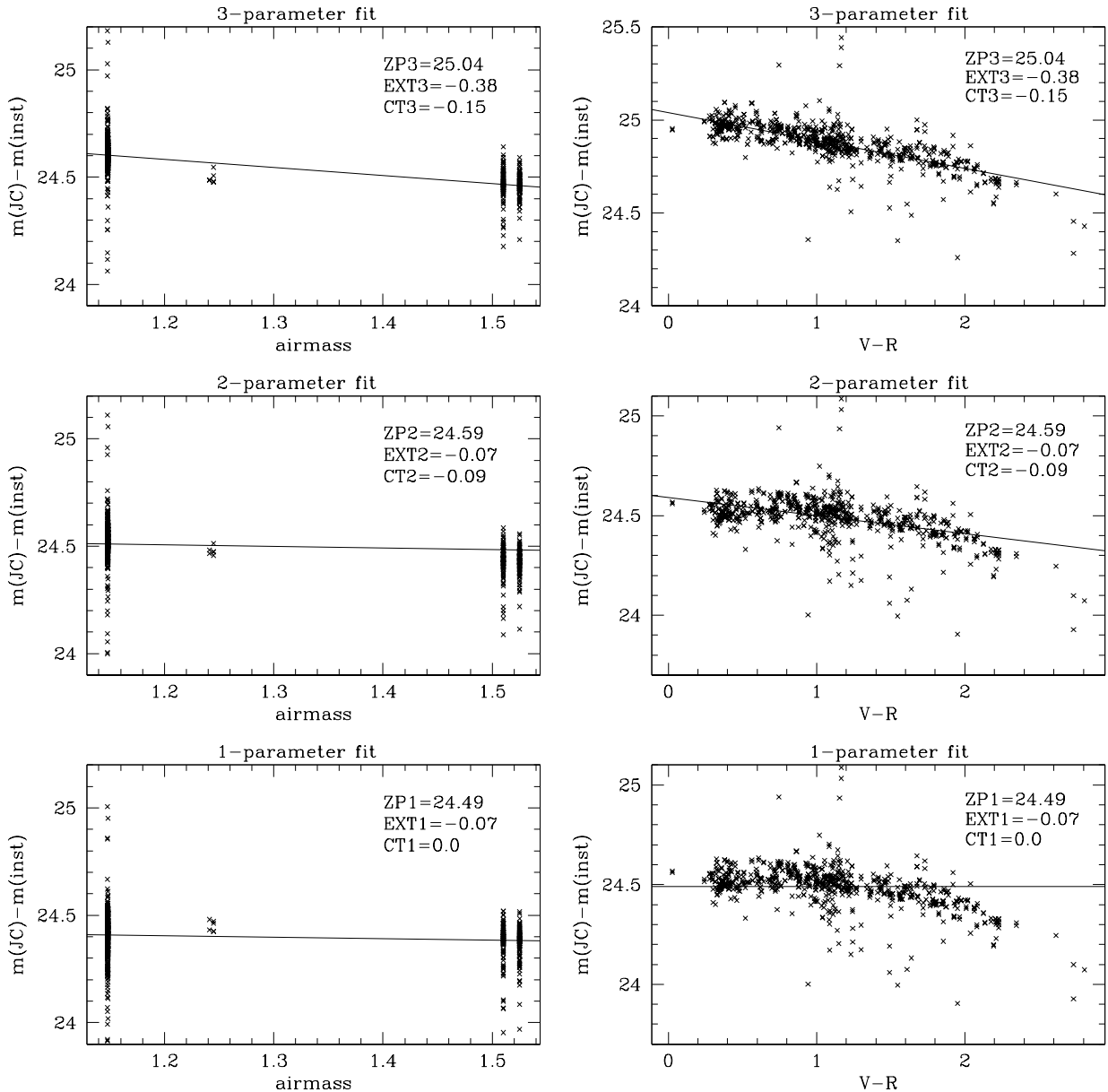}}
\caption{\label{fig:phot_solutions}Plot for the decision between
  photometric solutions of one night. The first row shows the results
  for the three-parameter fit, the difference between instrumental
  magnitude (data from all eight chips) and standard star magnitude
  plotted versus airmass (for colour zero) on the \emph{left} and
  versus colour (for airmass one) on the \emph{right}. In the second
  and third row the same diagrams are shown for the two- and the
  one-parameter fit. For this particular night, the three-parameter
  fit yields an extinction coefficient that is much too large
  (compared to the default value EXT$=-0.07$) resulting in a bad
  zeropoint. Obviously this night is not totally photometric.
  Nevertheless, experience shows that such nights can often be still
  used to estimate a zeropoint and sometimes even a colour term for
  the images. The one-parameter fit was chosen here because the
  non-linearities visible in the colour plots for the one- and the
  two-parameter fits influence the value for CT2, resulting again in a
  zeropoint ZP2 that is too large. For ideal photometric nights the
  values from the three different fits show only a very small scatter.
  The instrumental photometric errors are not shown in these plots
  since at the bright magnitudes of the Stetson/Landolt standard stars
  they are negligible and would hardly be visible.}
\end{figure*}

Here we describe in a bit more detail the absolute photometric
calibration which is not covered in \citet{2005astro.ph..1144E}.  All
objects from the standard star frames were extracted and matched to a
photometric standard star catalogue, the Landolt catalogue
\citep{1992AJ....104..340L} in the $U$-band and the Stetson catalogue
\citep{2000PASP..112..925S}\footnote{available at
  \texttt{http://cadcwww.hia.nrc.ca/cadcbin/\\wdbi.cgi/astrocat/stetson/query}}
for the other bands, respectively. Per WFI field, this usually led to
more than a thousand matched stars per night in the $BVRI$-bands and
to around a hundred in the $U$-band which could be used for
calibration.  The observed instrumental WFI magnitudes,
$m_{\mathrm{inst.}}$, were related to the standard Johnson-Cousins
system, $m_{\mathrm{JC}}$, by the following equation:

\begin{equation}
m_{\mathrm{JC}}=\mathrm{ZP}+m_{\mathrm{inst.}}+\mathrm{Colour}\cdot\mathrm{CT}+\mathrm{Airmass}\cdot\mathrm{EXT}\, ,
\end{equation}
with ZP the photometric zeropoint, CT the colour term, and EXT the
extinction coefficient. It should be stressed that the photometric
system of WFI is very different from the standard system and this
linear relation fails for some filters (especially the $U$- and
$B$-filters) and objects with large colour terms. It is in general
necessary to work with instrumental magnitudes when doing photometry
with WFI. Depending on the number of matched standard stars, different
solutions were chosen for the different nights. In nights with
standard star fields spanning over a wide range of airmasses, the
instrumental magnitudes could be fit to the standard system's
magnitudes with a three-parameter fit using the zeropoint, colour
term, and extinction coefficient as free parameters (for EXT only
negative values are fitted). For nights with poorer coverage in
airmass, the extinction term was fixed to a default value (taken from
the WFI website,
\texttt{http://www.ls.eso.org/lasilla/sciops/2p2/E2p2M/\\WFI/}) and a
two-parameter fit with the zeropoint and colour term as free
parameters was applied. Sometimes, especially in the $U$-band, also
colour coverage was not good enough to fit for the colour term. Then
only the zeropoint is estimated in a one-parameter fit with the colour
term also set to a default value. The decision was taken with the help
of plots showing the differences between instrumental magnitude and
standard star magnitude plotted versus airmass and versus colour (see
Fig.~\ref{fig:phot_solutions} for an example). The choice from these
plots should be regarded as the first run of our absolute photometric
calibration. Experience shows that it is often not possible to
entirely judge the photometric quality of a whole night from these
plots alone. Sometimes photometric conditions change over the night
which cannot be detected from these standard star exposures typically
observed two or three times a night only. Thus, we evaluate the
calibration by three further means, namely corrected photometric
zeropoints (see Sect.~\ref{sec:ast_phot_calib}),
colour-colour-diagrams of stars (see Fig.~\ref{fig:stellar_track}),
and apparent magnitude number-counts (see
Fig.~\ref{fig:numbercounts}). The method based on corrected
photometric zeropoints fully exploits the long term characteristic of
the DPS with observations in one field and colour usually spanning
many nights. All photometric solutions were added to the image header
so that later changes could be made (e.g.  from 2-parameter fit to
1-parameter fit).  Every night was given a unique so-called GaBoDS ID
starting with 1 on January, 1st, 1999.  The photometric solutions for
all the nights of the DPS can be found in the Appendix (see
Tables~\ref{tab:nights_U_35060} to \ref{tab:nights_I_EIS}). On
December, 13, 2004 the EIS team also published their photometric
solutions for all nights calibrated by them which can be found on the
web.\footnote{\texttt{http://www.eso.org/science/eis/surveys/release\_\\ 70000027\_Photometry.html}}

\subsection{Transition from run to set}
For every run websites were created containing essential image
information like exposure time, seeing, and other important
quantities.\footnote{available at:
  \url{http://marvin.astro.uni-bonn.de/DPS/}} The further steps like
relative astrometric and photometric calibration and coaddition
require all images from the same coordinates. Hence the images were
re-distributed from the runs into so-called sets containing
pre-reduced images from different epochs but at the same sky position.

\subsection{Astrometric and photometric calibration}
\label{sec:ast_phot_calib}
\begin{figure}
\resizebox{\hsize}{!}{\includegraphics{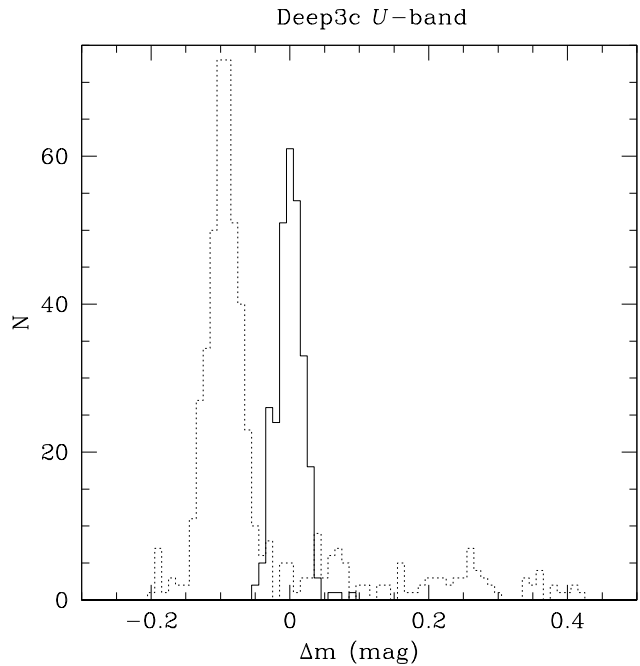}}
\caption{\label{fig:corr_ZP}Distribution of the differences $\Delta
  \mathrm{m}$ between the mean of the corrected zeropoints
  $\frac{1}{N}\sum_i \mathrm{ZP}_{\mathrm{corr},i}$ and the individual
  corrected zeropoints $\mathrm{ZP}_{\mathrm{corr},i}$ for the field
  Deep3c ($U$-band). The dashed line represents the distribution
  before selecting different photometric solutions by hand, and the
  solid line represents the distribution after this selection.}
\end{figure}

The astrometric calibration was performed with \emph{ASTROMETRIX}
\citep{2002ASTROMETRIX} on catalogues created with \emph{SExtractor}
\citep{1996A&AS..117..393B}. The external astrometry was fixed with
respect to the USNO-A2.0 \citep{1998yCat.1252....0M} catalogue. LDAC
tools were used to do the relative photometric calibration and to
bring all images to the same flux scale. First, the relative
zeropoints $\mathrm{ZP}_{\mathrm{rel}}$ of each single chip was
estimated from the flux differences of overlap objects. At this stage
it was required that the sum of the relative zeropoints of all images
of the set equals zero, $\sum_{i}^{}\,\mathrm{ZP}_{\mathrm{rel},i}=0$.
Then the images from calibrated nights (see Sect.~\ref{sec:abs_phot})
were taken and so-called corrected zeropoints
$\mathrm{ZP}_{\mathrm{corr}}$ were calculated in the following way:
\begin{equation}
\mathrm{ZP}_{\mathrm{corr},i}=\mathrm{ZP}+\mathrm{Airmass}\cdot\mathrm{EXT}+\mathrm{ZP}_{\mathrm{rel},i}\,
,
\end{equation}
with $\mathrm{ZP}$ and $\mathrm{EXT}$ being the zeropoint and the
extinction coefficient of the particular night, respectively.
Theoretically, these corrected zeropoints should then coincide for
photometric frames, and usually this assumption holds for most sets to
within $\sim0.05\,\mathrm{mag}$. Deviations were used as hints for bad
absolute photometric calibrations or changing photometric conditions
over the night. That is to say, if conditions change from the standard
star exposure to the scientific exposure over the course of the night,
this can be detected from the corrected zeropoints. Furthermore, it
can happen that from the plots shown in Fig.~\ref{fig:phot_solutions}
it is difficult to decide whether solution number two or three should
be chosen.  Sometimes it is even not entirely clear whether a night
should be rejected completely for the absolute photometric
calibration. At this point the distribution of the corrected
zeropoints can help to identify nights where obviously a bad decision
was taken. In Fig.~\ref{fig:corr_ZP} such a situation is illustrated
by an extreme example. There, the solutions chosen in the first run of
the absolute photometric calibration yielded a large scatter in the
corrected zeropoints. After rejecting some nights with small numbers
of standard stars or zeropoints far from the default values (hints for
non-photometric conditions) in the second run, the scatter in the
corrected zeropoints decreased considerably and the mean of this
distribution was taken as the zeropoint for the coadded image. After
such a treatment the distribution of the corrected photometric
zeropoints shows a HWHM scatter of $\sigma\la0.05\mathrm{mag}$ for all
images.

With the current calibration plan for WFI, taking standard star frames
mostly at the beginning and the end of the night (sometimes in
twilight), it is not possible to account for changing photometric
conditions during one night. Thus, a night has to be accepted or
rejected and no discrimination between good and bad data from one
night is feasible since the scientific data usually span the whole
night.

After the absolute photometric calibration, the fluxes of the images
were multiplied by a factor $10^{-0.4 ZP_{\mathrm{rel},i}}/t_i$,
with $t_i$ being the exposure time of the single image. Thus, the
counts in our final coadded image are scaled to one second.

\subsection{Coaddition}
The calibrated images then entered the coaddition which was performed
with \emph{SWarp} \citep{2003SWarp}. We chose a weighted mean
coaddition to maximise the $S/N$ of our final images relying on the
efficiency of our weight maps. Some quantities like seeing, sky
background, mean stellar ellipticities etc. were estimated for every
single frame during the relative astrometric and photometric
calibration step. By these values, single frames could be excluded
from the coaddition to avoid a degradation of the final image. In
fact, different coaddition conditions were often chosen so that for
some fields there is more than one image available in some filters.
For example, it can be useful for certain scientific purposes to have
an image as deep as possible (e.g. for deep multicolour photometry)
while seeing is not so important, whereas for other scientific
applications a very good seeing is mandatory (e.g. for weak lensing
studies).  The images were given a unique coaddition ID and the
selection criteria applied to the single frames entering the
coaddition were written to the FITS header (see
Table~\ref{tab:header}).

\section{Released data}
\label{sec:released-data}
\subsection{Images}

During the reduction of the DPS we produced 63 coadded images and
their corresponding weight maps. The basic properties of the coadded
images are summarised in Table~\ref{tab:DPS_images}. Usually the first
coaddition aimed at including as many single exposures as possible to
maximise the total exposure time. Exposures with a seeing of
$>2\arcsec$, unusual high background fluxes, or large relative
photometric zeropoints offsets compared to the rest of the set
(indicative of twilight, moonlight, or clouds) were excluded from the
coaddition. These coadded images were then assigned the letter ``A''
for ``All''. For example, the first coaddition of a Deep1c image then
has the coaddition ID D1CA. If the exposures allowed for a second
image with better seeing (but certainly with reduced exposure time) a
second coaddition was performed and the letter ``S'' for ``Seeing''
was assigned to this image (e.g.  coaddition ID D1BS). Sometimes, even
further conditions were applied to the single frames for another
coaddition then denoted by arbitrary letters like ``C'' or ``G''.
During 2005 more data from the GOODS program became available for the
field Deep2c. Thus, further images denoted by a letter ``B'' were
created with the same conditions as the images with assigned letter
``A'' but including more data. Only these ``D2CB'' images are shown in
Table~\ref{tab:DPS_images}.

The mispointings were processed, too. Therefore our data release
contains two new fields called Deep1e and Deep1f that are not original
DPS fields (see Table~\ref{tab:pos}) but that are nevertheless quite
deep in some bands and may be useful for some applications.

A description of the GaBoDS image headers can be found in
Appendix~\ref{sec:header}.

\subsection{Photometric accuracy}
\begin{figure}
\resizebox{\hsize}{!}{\includegraphics{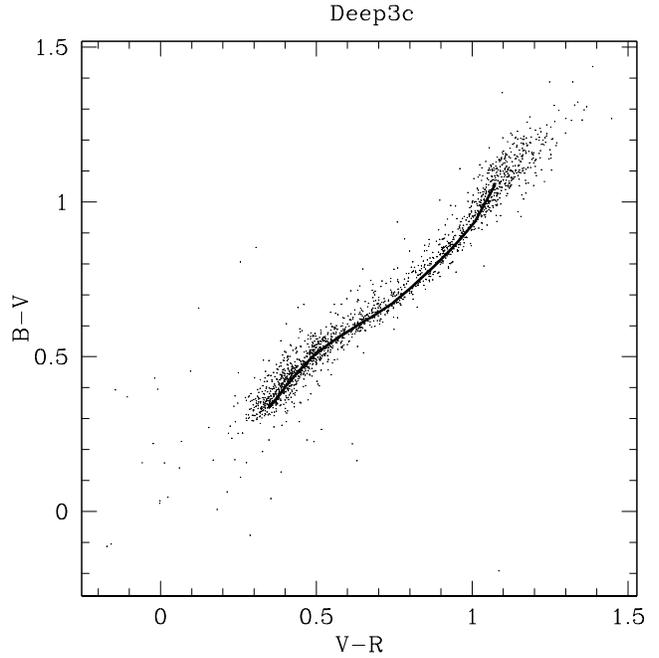}}
\caption{\label{fig:stellar_track} Colour-colour diagram for stars in
  the field Deep3c in comparison to an isochrone from
  \cite{2002A&A...391..195G}. The isochrone was calculated for stars
  with a metallicity of $Z=0.004$ ($\approx1/5\times$ solar) and an
  age of $10\,\mathrm{Gyr}$. Only stars with an initial mass of
  $m_{\mathrm{ini}}<0.92\,M_{\sun}$ were included.}
\end{figure}

Since we did not perform an illumination correction in the
pre-reduction step, a minimum photometric error of about
$0.05\,\mathrm{mag}$ \citep[see][]{2004AN....325..299K} is present
when using the same zeropoint over the whole field.  In order to check
the absolute photometric calibration, catalogues of all objects in the
coadded images were created and the colours of stars (selected by the
\emph{SExtractor}-CLASS\_STAR parameter) were plotted in colour-colour
diagrams. These measured colours were compared to theoretical stellar
isochrones by \cite{2002A&A...391..195G}. An exemplary plot is shown
in Fig.~\ref{fig:stellar_track}. After the thorough inspection of the
corrected zeropoints no offsets larger than $\sim0.15\mathrm{mag}$
(perpendicular to the track) were present in these colour-colour
diagrams.

Judging from these plots we estimate the error of our absolute
photometric calibration to be below $\sim0.1\,\mathrm{mag}$ for our
released images, in most cases better.

\begin{figure}
\resizebox{\hsize}{!}{\includegraphics{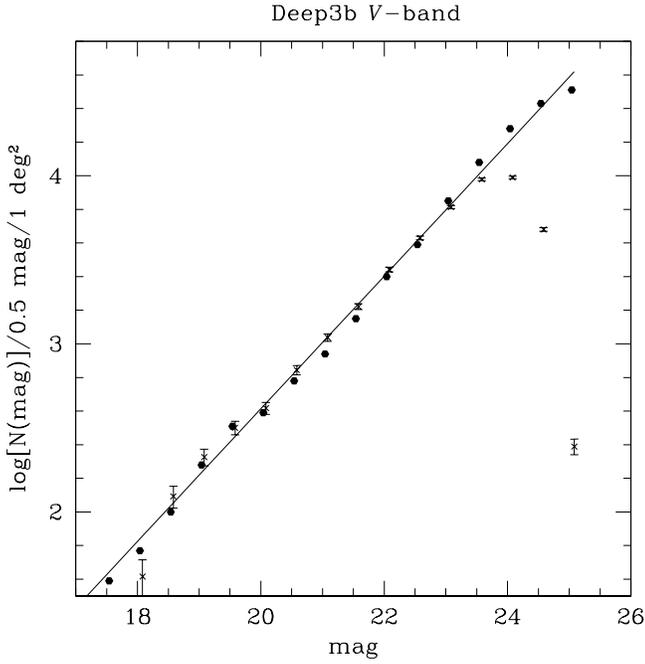}}
\caption{\label{fig:numbercounts}Apparent $V$-band magnitude number-counts for
  objects in the field Deep3b (crosses) in comparison to the well-calibrated
  field Deep2c (hexagons) showing no significant offset.}
\end{figure}

For an alternative check of the absolute photometric calibration, the
number-counts of objects were compared to the object number-counts of
our old, photometrically well-calibrated Chandra Deep Field South data
(ESO Press Photos 02a-d/03) in $BVR$ and to the Chandra Deep Field
South data from \citet{2001A&A...379..740A} in $U\_35060$, $U\_38$,
and $I$, respectively. An example is shown in
Fig.~\ref{fig:numbercounts}.  There is good agreement in these plots
for all images within the expected field-to-field variance (especially
at brighter magnitudes). For the $I\_EIS$ filter no such comparison
was performed.

\begin{figure}
\resizebox{\hsize}{!}{\includegraphics{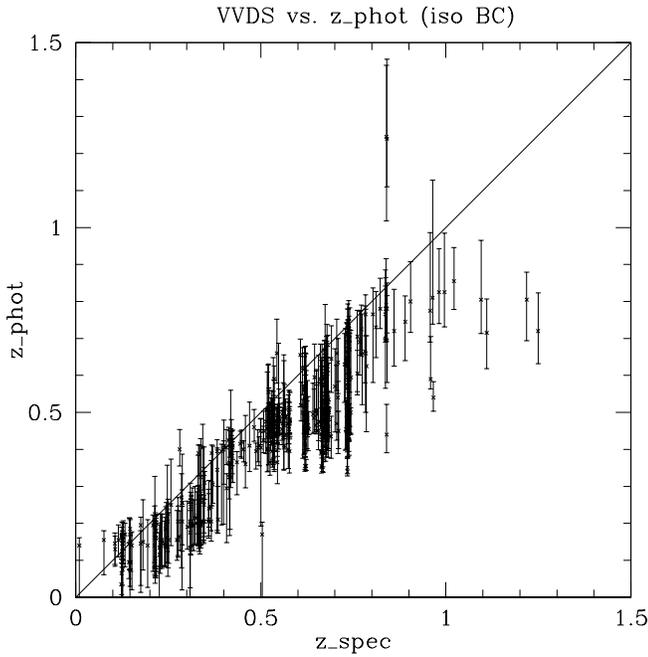}}
\caption{\label{fig:phot-z}Spectroscopic redshifts versus photometric
  redshifts for galaxies in the field Deep2c with $R<23$ from the VVDS
  (407 objects with secure spectroscopic redshifts and
  \emph{Hyperz}-68\%-confidence-intervals smaller than $0.23/(1+z)$).
  The errorbars represent the \emph{Hyperz}-68\%-confidence-interval.
  The standard deviation and the outlier fraction (see text) are
  $\sigma=0.14$ and $l=0.25\%$.  The systematic underestimation of the
  redshifts for galaxies with $z<1$ is reported in
  \cite{2000A&A...363..476B} for the $UBVRI$ filter set.}
\end{figure}

For the Chandra Deep Field South, an extensive database of
spectroscopically observed objects from the VIMOS-VLT-Deep-Survey
\citep[VVDS, ][]{2004A&A...428.1043L} exists. We compared the
spectroscopic redshifts of all 407 VVDS galaxies with $R<23$ and
secure spectroscopic redshifts to photometric redshift estimates from
the public photometric redshift code \emph{Hyperz}
\citep{2000A&A...363..476B} based on our $UBVRI$ photometry in the
field Deep2c. Therefore we matched the seeing of the images in the
different bands by convolving them with Gaussian kernels of the
appropriate size. For the photometric redshift estimation we used the
templates created from the synthetic stellar libraries of
\cite{1993ApJ...405..538B}. The results are shown in
Fig.~\ref{fig:phot-z}. An outlier is defined as an object with
$\left|z_{\mathrm{spec}}-z_{\mathrm{phot}}\right|>1$ and the
dispersion is estimated via
\[\sigma^2=\frac{\sum_{i=1}^{N}\left(z_{\mathrm{spec},i}-z_{\mathrm{phot},i}\right)^2}{N}\,
,\] excluding outliers. These results are at least as good as the
estimates from simulations given in \cite{2000A&A...363..476B} for the
$UBVRI$ filter set, thus indicating a robust photometric calibration.

\subsection{Astrometric accuracy}
\label{sec:astrometric_accuracy}
\begin{figure}
\resizebox{\hsize}{!}{\includegraphics{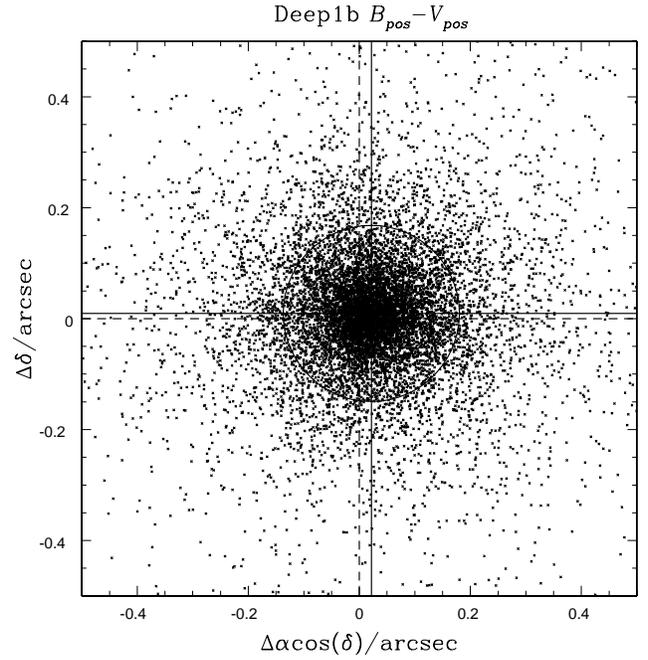}}
\caption{\label{fig:diff_pos} Distribution of the positional
  differences of objects in the Deep1b $B$- and $V$-band images. The
  mean and standard deviations here are:
  $\langle{\Delta\alpha\mathrm{cos}(\delta)}\rangle=0\farcs02 \pm
  0\farcs16$ and $\langle{\Delta{\delta}}\rangle=0\farcs01 \pm
  0\farcs15$. The solid lines mark the centre and the circle
  represents the one sigma interval (radius: $0\farcs15$).}
\end{figure}

\begin{figure}
\resizebox{\hsize}{!}{\includegraphics{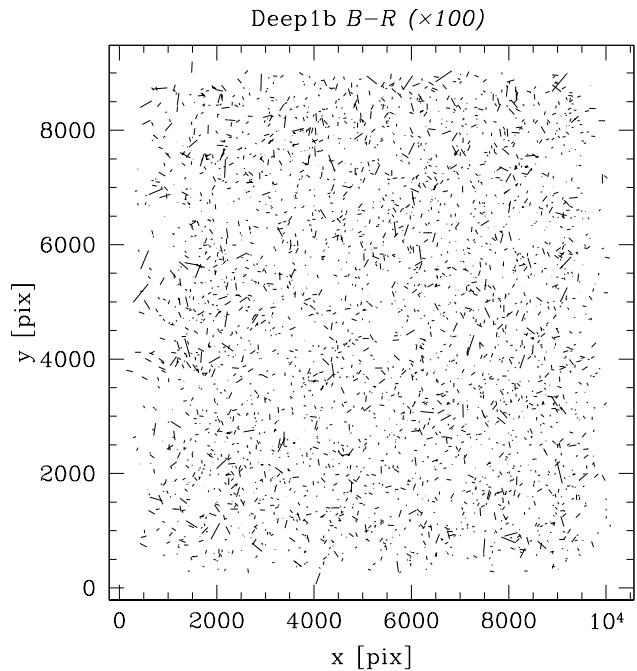}}
\caption{\label{fig:sticks_B_V} Dependence of the positional differences on
  chip position. The distribution seems to be random and no chip boundaries
  etc. are visible. The differences have been multiplied by a factor 100 and
  the longest sticks correspond to $\sim1\farcs5$.}
\end{figure}

While overlap astrometry was used for all single chips entering the
coaddition in one band, the final images of different bands were
processed astrometrically independently and the catalogues produced
for the colour-colour diagrams are also used to check the internal
astrometric accuracy between the different colours. The positional
differences of associated objects from the $B$- and $V$-band images
for the field Deep1b are plotted in Fig.~\ref{fig:diff_pos}. The
dependence of these differences on chip position is shown in
Fig~\ref{fig:sticks_B_V} where no residual effects like chip
boundaries etc. are visible. For all fields, the distribution of the
position differences between two bands are well described by a
Gaussian with a half width at half maximum of $\sigma\la0\farcs2$ (see
Table~\ref{tab:DPS_images}). A slightly higher internal astrometric
accuracy could have been obtained by using an astrometric catalogue
from one band as the reference for the other bands. Given our already
high accuracy of $\sigma< 1\mathrm{pix}$ this step was not performed
for flexibility reasons.  The quoted accuracy should suffice for
almost all scientific goals.

The external astrometric accuracy is checked with the help of the
UCAC-2 catalogue \citep{2004AJ....127.3043Z}. No offsets larger than
the accuracy of the USNO-A2.0 catalogue of $0\farcs3$ are found.

\subsection{Websites}
The released images are accompanied by websites which give quick
access to a lot of information about the data reduction. The pages are
organised as follows.  At the homepage
(\url{http://marvin.astro.uni-bonn.de/DPS/}) links lead to pages for
every filter with further links to all observing runs in this band.
These run webpages consist of pages for the calibration frames (BIAS,
DARK, SKYFLAT) and for the scientific frames. For the calibration
images, thumbnails for visual inspection of the master calibration
frame and lists for all single frames used for the creation of the
master image are presented. For the scientific frames, basic
statistics are listed and a thumbnail of every single frame can be
inspected together with a plot of the measured PSF. Further links lead
to the photometric solutions of the actual night with plots used for
the decision between the different photometric solutions (see
Fig.~\ref{fig:phot_solutions}).

For every set, websites listing all coadded images in a particular
band and their basic properties are also linked from the homepage.
Plots showing the distribution of single exposure quantities like
seeing, relative photometric zeropoint, etc. are featured as well as
colour-colour-plots for all possible combinations. The website for a
coadded image contains crosslinks to the run websites of the single
frames included and a number of checkplots created after coaddition.
In particular, there are checkplots showing the number-counts (see
Fig.~\ref{fig:numbercounts}) and angular correlation functions of
galaxies in different magnitude
bins.% (see Fig.~\ref{fig:correlation}).

%\begin{figure}
%\resizebox{\hsize}{!}{\includegraphics{DPS_images/correlation.pdf}}
%\caption{\label{fig:correlation}Angular correlation function for galaxies in
%  the Deep1b $I$-band image in two different magnitude ranges.}
%\end{figure}

\section{Comparison to the EIS DPS optical data release}
\label{sec:EIS_compare}
\subsection{Released data}
On November 10, 2004 the EIS team released 40 reduced and stacked
optical images of the DPS to the scientific
community.\footnote{available
  at:\texttt{http://www.eso.org/science/eis/surveys/\\release\_60000024\_DPS.html}}
Their reduction includes data accumulated until September 28, 2002.
Compared to our data release, there are no images for the
mispointings, for Deep1a in the $V$-band, for Deep1c in the
$I\_EIS$-band, for Deep2c in the $B$- and $R$-bands, for Deep3a in the
$I$-band, and for Deep3d in the $B$- and $R$-bands.  Furthermore, we
used more data for the fields Deep1b ($U$), Deep2b ($I$), Deep2c (all
bands), and Deep3b ($U$), reaching greater depth.  The EIS release
includes an image of the field Deep2c in the $U\_38$-band which is not
included in our release.

\subsection{Photometric comparison}
After some comparisons between objects' magnitudes in our reduction
and in the EIS data release it became clear that the absolute
photometric calibration of the EIS images is somewhat different from
our calibration. Absolute offsets of some tenths of a magnitude are
present for most images. These offsets are also present in
colour-colour diagrams when the EIS data are compared to theoretical
isochrones by \citet{2002A&A...391..195G}. Intense cooperation with
the EIS team led to the discovery of a mistake in their final
application of the absolute photometric calibration to the images
(Miralles, private communication) so that no further comparison is
meaningful here.

Nevertheless, if both pipelines performed a correct relative
photometry between the eight WFI chips there should be only a constant
offset between the magnitudes of objects measured in an EIS image and
the same objects measured in one of our images. Within photometric
errors this is indeed the case for all images from the Deep1 and Deep3
regions. For images from the Deep2 region, the scatter of the
magnitude differences is much larger. A careful cooperative analysis
with EIS revealed another mistake in the EIS reduction (Miralles,
private communication).  The chips were not brought to a common
zeropoint during the reduction of the Deep2 fields in contrast to
Deep1 and Deep3.

\subsection{Astrometric comparison}
The EIS team used the GSC2.2 catalogue for the absolute astrometric
calibration which is known to be offset with respect to the USNO-A2.0
catalogue we used. We do not find any offsets between our images and
the EIS images of larger than $\sim0\farcs5$. The actual offset
depends strongly on the sky position with the best match in the Deep1
fields and the worst in the Deep3 fields. The scatter is comparable to
the scatter between the different colour catalogues from our images
(see Sect.~\ref{sec:astrometric_accuracy} and
Table~\ref{tab:DPS_images}).

Table~\ref{tab:compare_EIS} summarises the comparisons between our
data release and the EIS data.

\section{Summary}
\label{sec:summary}
We release 63 fully reduced and stacked images of the ESO Deep Public
Survey to the scientific community. These images, reduced with our
THELI pipeline, are shown to have good internal and external
astrometric and photometric calibrations. A comparison to the EIS-DPS
release shows significant offsets in the absolute photometric
calibration for most images and large scatter in the magnitude
differences of objects for a subset. Both effects are shown to
originate from the EIS data and this is acknowledged by the EIS team.

The optical data from the DPS are currently used by our group to do searches
for Lyman-break galaxies. \citet{2005A&A....Hildebrandt} have selected large
samples of $U$- and $B$-dropouts in deep WFI images of the CDFS. The methods
presented there will be applied to the whole DPS dataset.

\begin{acknowledgements}
  This work was supported by the German Ministry for Education and
  Science (BMBF) through the DLR under the project 50 OR 0106, by the
  BMBF through DESY under the project 05AE2PDA/8, and by the Deutsche
  Forschungsgemeinschaft (DFG) under the projects SCHN342/3-1 and
  ER327/2-1.
  
  We would like to thank Dr. L. Girardi very much for creating additional
  isochrones for the new WFI filters.

  Many thanks also to the EIS team for fruitful discussions and
  comparisons of the two different reductions.
\end{acknowledgements}

\bibliographystyle{aa}

\bibliography{4278}

\begin{longtable}{l l l r c r c r r}
  \caption{\label{tab:DPS_images}Released optical images from the DPS reduced with the THELI pipeline. The first three characters of the coaddition ID in column three represent the field (D1A for field Deep1a etc.) and the fourth character is an identifier to distinguish the different coadditions of one field. The Vega limiting magnitudes in column six correspond to  $5\sigma$ sky level measured in a circular aperture of $2\arcsec$ radius. The grades in column seven are estimated from visual inspection of the coadded images. Grade ``A'' is assigned to an image with no special features and an appearance typical for the particular band. Grade ``B'' is assigned to an image with cosmetic defects. Nevertheless also the grade ``B'' images are fully usable for most scientific purposes. Astrometric offsets in columns eight and nine are given with respect to the $R$-band image (coadd ID D??A and D2CB for the field Deep2c, respectively) of that field.}\\
  \hline \hline
  Field & Filter & coadd ID & exp. time & FWHM        & mag lim. & Grade& $\Delta\alpha\mathrm{cos}(\delta)$ & $\Delta\delta$\\
  &        &          & [s]       & [$\arcsec$] & [mag]& & [$\arcsec$] & [$\arcsec$]  \\
  \hline \endfirsthead
\caption{Released optical images from the DPS reduced with the THELI
  pipeline.}\\
\hline\hline
Field & Filter & coadd ID & exp. time & FWHM        & mag lim. & Grade & $\Delta\alpha\mathrm{cos}(\delta)$ & $\Delta\delta$\\
      &        &          & [s]       & [$\arcsec$] & [mag]& & [$\arcsec$] & [$\arcsec$]  \\
\hline \endhead
\endfoot
Deep1a & $U\_35060$ & D1AA & $53\,095$ & 1.34 & 25.45 & A & $-0.01\pm0.16$ & $-0.01\pm0.16$ \\ 
Deep1a & $U\_35060$ & D1AC & $71\,093$ & 1.44 & 25.57 & A & $-0.01\pm0.15$ & $-0.01\pm0.15$ \\ 
Deep1a & $U\_38$ & D1AA & $17\,398$ & 1.28 & 24.46 & A & $-0.00\pm0.14$ & $-0.01\pm0.14$ \\ 
Deep1a & $B$ & D1AA & $11\,696$ & 1.34 & 26.38 & A & $-0.00\pm0.14$ & $-0.01\pm0.14$ \\ 
Deep1a & $V$ & D1AA & $8684$ & 0.98 & 25.76 & A & $0.01\pm0.12$ & $-0.02\pm0.12$ \\ 
Deep1a & $R$ & D1AA & $9597$ & 0.85 & 25.50 & A & - & - \\ 
Deep1a & $I$ & D1AA & $31\,493$ & 0.84 & 24.33 & A & $0.00\pm0.12$ & $-0.00\pm0.11$ \\ 
Deep1a & $I$ & D1AG & $25\,194$ & 0.84 & 24.29 & A & $0.00\pm0.12$ & $-0.00\pm0.11$ \\ 
\hline
Deep1b & $U\_35060$ & D1BA & $46\,793$ & 1.06 & 25.27 & A & $-0.03\pm0.15$ & $-0.01\pm0.15$ \\ 
Deep1b & $B$ & D1BA & $9597$ & 1.33 & 26.09 & A & $-0.02\pm0.15$ & $-0.01\pm0.14$ \\ 
Deep1b & $B$ & D1BS & $4199$ & 1.09 & 25.61 & A & $-0.02\pm0.13$ & $-0.00\pm0.13$ \\ 
Deep1b & $V$ & D1BA & $9897$ & 1.32 & 25.60 & A & $-0.00\pm0.14$ & $0.00\pm0.13$ \\ 
Deep1b & $V$ & D1BS & $4499$ & 1.20 & 25.24 & A & $0.00\pm0.13$ & $-0.00\pm0.12$ \\ 
Deep1b & $R$ & D1BA & $19\,794$ & 1.28 & 25.70 & B & - & - \\ 
Deep1b & $R$ & D1BS & $8398$ & 1.10 & 25.28 & A & $0.00\pm0.10$ & $0.00\pm0.09$ \\ 
Deep1b & $I$ & D1BA & $25\,493$ & 0.97 & 24.38 & A & $-0.01\pm0.13$ & $-0.00\pm0.12$ \\ 
\hline
Deep1c & $V$ & D1CA & $7498$ & 1.19 & 25.47 & A & $-0.03\pm0.16$ & $-0.00\pm0.15$ \\ 
Deep1c & $V$ & D1CS & $4199$ & 1.11 & 25.25 & A & $-0.03\pm0.15$ & $-0.00\pm0.14$ \\ 
Deep1c & $R$ & D1CA & $11\,696$ & 0.97 & 25.28 & A & - & - \\ 
Deep1c & $I\_EIS$ & D1CA & $4445$ & 1.25 & 23.34 & A & $-0.04\pm0.15$ & $-0.02\pm0.14$ \\ 
\hline
Deep1e & $U\_38$ & D1EA & $9899$ & 1.78 & 24.34 & A & $-0.01\pm0.09$ & $-0.03\pm0.21$ \\ 
Deep1e & $R$ & D1EA & $8998$ & 0.90 & 25.37 & A & - & - \\ 
Deep1e & $I$ & D1EA & $11\,398$ & 1.26 & 23.93 & A & $-0.01\pm0.07$ & $-0.02\pm0.15$ \\ 
\hline
Deep1f & $I$ & D1FA & $14\,997$ & 1.16 & 23.92 & A & - & - \\ 
\hline
Deep2a & $R$ & D2AA & $5998$ & 0.83 & 25.11 & A & - & - \\ 
\hline
Deep2b & $U\_35060$ & D2BA & $53\,095$ & 1.15 & 25.23 & A & $-0.02\pm0.14$ & $-0.01\pm0.14$ \\ 
Deep2b & $U\_35060$ & D2BC & $64\,794$ & 1.27 & 25.30 & A & $-0.02\pm0.14$ & $-0.01\pm0.14$ \\ 
Deep2b & $B$ & D2BA & $11\,396$ & 0.98 & 26.23 & A & $0.00\pm0.13$ & $0.00\pm0.13$ \\ 
Deep2b & $B$ & D2BS & $9597$ & 0.91 & 26.13 & A & $0.00\pm0.13$ & $0.00\pm0.13$ \\ 
Deep2b & $V$ & D2BA & $9297$ & 0.88 & 25.59 & A & $-0.00\pm0.12$ & $0.01\pm0.12$ \\ 
Deep2b & $R$ & D2BA & $10\,497$ & 1.33 & 25.42 & A & - & - \\ 
Deep2b & $R$ & D2BS & $4799$ & 1.09 & 25.01 & A & $0.00\pm0.08$ & $0.00\pm0.08$ \\ 
Deep2b & $I$ & D2BA & $19\,196$ & 0.75 & 24.25 & A & $0.00\pm0.12$ & $0.01\pm0.11$ \\ 
\hline
Deep2c & $U\_35060$ & D2CB & $78\,891$ & 1.01 & 25.49 & A & $-0.02\pm0.13$ & $0.03\pm0.14$ \\ 
Deep2c & $B$ & D2CA & $69\,431$ & 0.98 & 27.28 & A & $0.01\pm0.11$ & $-0.00\pm0.11$ \\ 
Deep2c & $V$ & D2CB & $104\,603$ & 0.92 & 26.77 & A & $0.01\pm0.10$ & $0.00\pm0.10$ \\ 
Deep2c & $R$ & D2CB & $87\,654$ & 0.79 & 26.54 & A & - & - \\ 
Deep2c & $I$ & D2CA & $34\,575$ & 0.93 & 24.49 & A & $0.00\pm0.11$ & $0.00\pm0.11$ \\ 
\hline
Deep2d & $R$ & D2DA & $2999$ & 1.06 & 24.81 & A & - & - \\ 
\hline
Deep3a & $U\_35060$ & D3AA & $26\,997$ & 1.09 & 24.77 & A & $-0.01\pm0.13$ & $-0.01\pm0.13$ \\ 
Deep3a & $U\_35060$ & D3AC & $35\,996$ & 1.10 & 24.91 & A & $-0.01\pm0.13$ & $-0.00\pm0.13$ \\ 
Deep3a & $U\_38$ & D3AA & $26\,997$ & 1.27 & 24.59 & A & $-0.01\pm0.12$ & $-0.00\pm0.12$ \\ 
Deep3a & $B$ & D3AA & $11\,096$ & 0.92 & 26.08 & A & $-0.00\pm0.12$ & $-0.00\pm0.11$ \\ 
Deep3a & $V$ & D3AA & $8998$ & 1.02 & 25.44 & A & $-0.00\pm0.11$ & $-0.00\pm0.11$ \\ 
Deep3a & $R$ & D3AA & $8997$ & 0.81 & 25.28 & A & - & - \\ 
Deep3a & $R$ & D3AS & $7798$ & 0.79 & 25.23 & A & $-0.00\pm0.05$ & $0.00\pm0.05$ \\ 
Deep3a & $I$ & D3AA & $23\,396$ & 0.95 & 24.22 & B & $0.01\pm0.11$ & $0.00\pm0.11$ \\ 
Deep3a & $I$ & D3AG & $13\,198$ & 0.91 & 23.94 & A & $0.00\pm0.11$ & $0.00\pm0.11$ \\ 
\hline
Deep3b & $U\_35060$ & D3BA & $53\,995$ & 1.01 & 25.13 & A & $0.01\pm0.13$ & $-0.00\pm0.13$ \\ 
Deep3b & $B$ & D3BA & $9897$ & 0.95 & 26.06 & A & $0.00\pm0.12$ & $-0.00\pm0.12$ \\ 
Deep3b & $V$ & D3BA & $8998$ & 0.88 & 25.45 & A & $0.00\pm0.11$ & $0.00\pm0.11$ \\ 
Deep3b & $R$ & D3BA & $9297$ & 0.79 & 25.30 & A & - & - \\ 
Deep3b & $I$ & D3BA & $26\,993$ & 0.82 & 24.27 & A & $0.00\pm0.11$ & $0.01\pm0.10$ \\ 
\hline
Deep3c & $U\_35060$ & D3CA & $46\,845$ & 0.97 & 25.22 & A & $0.01\pm0.13$ & $-0.00\pm0.13$ \\ 
Deep3c & $B$ & D3CA & $13\,496$ & 0.93 & 26.24 & A & $0.01\pm0.12$ & $0.00\pm0.12$ \\ 
Deep3c & $B$ & D3CS & $11\,996$ & 0.90 & 26.17 & A & $0.01\pm0.12$ & $-0.00\pm0.12$ \\ 
Deep3c & $V$ & D3CA & $5998$ & 0.79 & 25.30 & A & $0.01\pm0.11$ & $-0.00\pm0.10$ \\ 
Deep3c & $R$ & D3CA & $8998$ & 0.81 & 25.26 & A & - & - \\ 
Deep3c & $I$ & D3CA & $25\,193$ & 1.02 & 24.04 & B & $0.01\pm0.11$ & $0.00\pm0.11$ \\ 
\hline
Deep3d & $B$ & D3DA & $11\,097$ & 0.88 & 26.22 & A & $0.01\pm0.12$ & $-0.01\pm0.12$ \\ 
Deep3d & $V$ & D3DA & $8998$ & 0.90 & 25.37 & A & $0.01\pm0.11$ & $-0.00\pm0.10$ \\ 
Deep3d & $R$ & D3DA & $8998$ & 0.73 & 24.76 & A & - & - \\ 
Deep3d & $I\_\mathrm{EIS}$ & D3DA & $22\,293$ & 0.76 & 24.51 & A & $0.01\pm0.10$ & $-0.00\pm0.10$ \\ 
\hline
\end{longtable}

\Online
\appendix
\section{Photometric calibrations}
In Tables~\ref{tab:nights_U_35060} to \ref{tab:nights_I_EIS} the
photometric solutions for all calibrated nights are shown. The
solution chosen for a particular coadded image can be found in the
FITS image header.
\begin{table*}
  \caption{\label{tab:nights_U_35060}
    Photometric solutions in the $U\_35060$-band. The colour term corresponds to the Johnson-Cousins colour $U-B$. The default value for the extinction coefficient in the one- and two-parameter fits is EXT1~$=$~EXT2~$=-0.48$ and the default value for the colour term in the one-parameter fit is CT1~$=0.05$.}
\begin{tabular}{c r|r r r|r r|r }
\hline \hline
night & GaBoDS ID & ZP3 & EXT3 & CT3 & ZP2 & CT2 & ZP1 \\
\hline
2000-10-26 & 665 & 20.48 & $-$0.00 & $-$10.04 & 21.15 & $-$10.21 & 22.06 \\
2000-10-27 & 666 & 22.11 & $-$0.47 & $-$0.11 & 22.12 & $-$0.11 & 22.12 \\
2000-11-27 & 697 & 23.81 & $-$1.93 & 0.02 & 22.13 & 0.02 & 22.11 \\
2000-11-28 & 698 & 27.06 & $-$4.64 & 0.05 & 22.12 & 0.05 & 22.12 \\
2001-02-21 & 783 & 21.21 & 0.00 & 0.02 & 21.79 & 0.02 & 21.77 \\
2001-02-22 & 784 & 21.35 & 0.00 & 0.09 & 21.95 & 0.10 & 21.97 \\
2001-02-23 & 785 & 21.78 & $-$0.26 & 0.10 & 22.07 & 0.10 & 22.09 \\
2001-02-26 & 788 & 21.85 & $-$0.37 & 0.08 & 21.99 & 0.08 & 22.00 \\
2001-03-24 & 814 & 22.19 & $-$0.62 & 0.07 & 22.02 & 0.07 & 22.03 \\
2001-03-25 & 815 & 22.55 & $-$0.94 & 0.09 & 21.99 & 0.09 & 22.00 \\
2001-04-20 & 841 & 23.09 & $-$1.38 & 0.10 & 21.78 & 0.10 & 21.80 \\
2001-04-21 & 842 & 21.88 & $-$0.36 & 0.07 & 22.04 & 0.06 & 22.04 \\
2001-07-20 & 932 & 25.73 & $-$3.23 & 0.23 & 22.04 & 0.22 & 22.17 \\
2001-07-21 & 933 & 23.36 & $-$1.51 & 0.37 & 21.92 & 0.44 & 22.11 \\
2001-07-22 & 934 & 22.82 & $-$0.87 & 0.14 & 22.33 & 0.13 & 22.38 \\
2001-07-23 & 935 & 21.58 & $-$0.00 & 0.11 & 22.21 & 0.11 & 22.26 \\
2001-07-24 & 936 & 22.24 & $-$0.40 & 0.19 & 22.34 & 0.19 & 22.41 \\
%2001-07-25 & 937 & 46.48 & $-$18.46 & 0.17 & 22.05 & 0.41 & 22.31 \\
2001-07-25 & 937 & - & - & - & 22.05 & 0.41 & 22.31 \\
%2001-07-26 & 938 & 53.89 & $-$24.84 & 0.52 & 21.92 & 0.52 & 22.21 \\
2001-07-26 & 938 & - & - & - & 21.92 & 0.52 & 22.21 \\
2001-08-19 & 962 & 22.46 & $-$0.54 & 0.08 & 22.37 & 0.08 & 22.38 \\
2001-08-20 & 963 & 22.40 & $-$0.51 & 0.09 & 22.37 & 0.09 & 22.39 \\
2001-08-21 & 964 & 22.90 & $-$0.91 & 0.04 & 22.37 & 0.05 & 22.37 \\
2001-08-22 & 965 & 26.52 & $-$3.95 & 0.02 & 22.35 & 0.05 & 22.35 \\
2001-08-23 & 966 & 22.35 & $-$0.48 & 0.06 & 22.35 & 0.06 & 22.35 \\
2001-11-13 & 1048 & 21.73 & 0.00 & 0.01 & 22.29 & 0.01 & 22.27 \\
2001-11-14 & 1049 & 22.97 & $-$1.06 & 0.08 & 22.29 & 0.09 & 22.31 \\
2001-11-15 & 1050 & 22.78 & $-$0.90 & 0.05 & 22.28 & 0.05 & 22.28 \\
2002-03-09 & 1164 & 22.10 & $-$0.47 & 0.04 & 22.12 & 0.04 & 22.12 \\
2002-03-10 & 1165 & 22.01 & $-$0.45 & 0.16 & 22.04 & 0.16 & 22.10 \\
2002-03-11 & 1166 & 21.95 & $-$0.28 & 0.11 & 22.20 & 0.12 & 22.23 \\
2002-03-12 & 1167 & 22.12 & $-$0.37 & 0.05 & 22.26 & 0.06 & 22.26 \\
2002-03-13 & 1168 & 21.73 & 0.00 & 0.03 & 22.28 & 0.03 & 22.27 \\
2002-06-07 & 1254 & 21.49 & 0.00 & 0.57 & 22.07 & 0.56 & 22.33 \\
2002-06-09 & 1256 & 22.88 & $-$1.00 & $-$0.11 & 22.24 & $-$0.08 & 22.23 \\
2002-06-10 & 1257 & 22.18 & $-$0.39 & 0.08 & 22.28 & 0.08 & 22.29 \\
%2002-06-12 & 1259 & 61.53 & $-$31.32 & 1.57 & 21.82 & 2.50 & 22.06 \\
2002-06-12 & 1259 & - & - & - & 21.82 & 2.50 & 22.06 \\
2002-08-11 & 1319 & 22.02 & $-$0.35 & $-$0.69 & 22.18 & $-$0.71 & 21.99 \\
%2002-09-28 & 1367 & 31.61 & $-$8.82 & 0.70 & 21.86 & 0.70 & 22.20 \\
2002-09-28 & 1367 & - & - & - & 21.86 & 0.70 & 22.20 \\
2002-12-11 & 1441 & 22.28 & $-$0.54 & 0.12 & 22.20 & 0.12 & 22.24 \\
2004-05-12 & 1959 & 21.58 & 0.00 & $-$0.14 & 22.15 & $-$0.14 & 22.12 \\
2004-10-31 & 2131 & 23.72 & $-$1.67 & 0.02 & 22.35 & 0.03 & 22.35 \\
2004-11-02 & 2133 & - & - & - & - & - & 22.32 \\
2005-10-30 & 2495 & 21.81 & $-$0.09 & 0.07 & 22.40 & 0.00 & 22.36 \\
\hline
\end{tabular}
\end{table*}

\begin{table*}
\caption{\label{tab:nights_U_38}
Photometric solutions in the $U\_38$-band. The colour term corresponds to the Johnson-Cousins colour $U-B$. The default value for the extinction coefficient in the one- and two-parameter fits is EXT1~$=$~EXT2~$=-0.73$ and the default value for the colour term in the one-parameter fit is CT1~$=-0.01$.}
\begin{tabular}{c r|r r r|r r|r }
\hline \hline
night & GaBoDS ID & ZP3 & EXT3 & CT3 & ZP2 & CT2 & ZP1 \\
\hline
2000-03-30 & 455 & 21.24 & 0.00 & 0.03 & 22.09 & 0.03 & 22.11 \\
2000-03-31 & 456 & 21.21 & 0.00 & 0.05 & 22.07 & 0.05 & 22.10 \\
2000-04-01 & 457 & 25.19 & $-$3.37 & 0.05 & 22.09 & 0.05 & 22.12 \\
2000-04-05 & 461 & 21.17 & 0.00 & 0.04 & 22.03 & 0.04 & 22.05 \\
2000-07-29 & 576 & - & - & - & 23.76 & 16.57 & 22.15 \\
2000-08-01 & 579 & - & - & - & 23.09 & 9.37 & 22.18 \\
2000-08-26 & 604 & 21.34 & $-$0.06 & 0.09 & 22.17 & 0.08 & 22.24 \\
2000-08-27 & 605 & 20.97 & 0.00 & 0.55 & 21.84 & 0.55 & 22.17 \\
\hline
\end{tabular}
\end{table*}

\begin{table*}
  \caption{\label{tab:nights_B}
    Photometric solutions in the $B$-band. The colour term corresponds to the Johnson-Cousins colour $B-V$. The default value for the extinction coefficient in the one- and two-parameter fits is EXT1~$=$~EXT2~$=-0.22$ and the default value for the colour term in the one-parameter fit is CT1~$=0.25$.}
\begin{tabular}{c r|r r r|r r|r }
\hline \hline
night & GaBoDS ID & ZP3 & EXT3 & CT3 & ZP2 & CT2 & ZP1 \\
\hline
%1999-12-02 & 336 & 38.52 & $-$13.35 & 0.27 & 24.74 & 0.27 & 24.76 \\
1999-12-02 & 336 & - & - & - & 24.74 & 0.27 & 24.76 \\
2000-03-29 & 454 & 25.61 & $-$0.99 & 0.27 & 24.59 & 0.26 & 24.60 \\
2000-03-30 & 455 & 24.72 & $-$0.29 & 0.23 & 24.63 & 0.23 & 24.61 \\
2000-03-31 & 456 & 24.83 & $-$0.38 & 0.23 & 24.63 & 0.23 & 24.62 \\
2000-04-01 & 457 & 24.40 & 0.00 & 0.21 & 24.66 & 0.21 & 24.63 \\
2000-04-06 & 462 & 24.43 & $-$0.00 & 0.20 & 24.69 & 0.20 & 24.65 \\
2000-07-02 & 549 & 26.29 & $-$1.53 & 0.17 & 24.77 & 0.18 & 24.69 \\
2000-11-27 & 697 & 24.42 & 0.00 & 0.22 & 24.70 & 0.22 & 24.68 \\
2000-11-28 & 698 & 24.71 & $-$0.23 & 0.22 & 24.69 & 0.22 & 24.67 \\
2001-02-01 & 763 & 24.66 & $-$0.23 & 0.23 & 24.65 & 0.23 & 24.63 \\
2001-02-02 & 764 & 24.45 & $-$0.16 & 0.23 & 24.53 & 0.23 & 24.51 \\
2001-02-25 & 787 & 24.66 & $-$0.27 & 0.23 & 24.58 & 0.23 & 24.57 \\
2001-02-26 & 788 & 24.51 & $-$0.14 & 0.22 & 24.61 & 0.22 & 24.59 \\
2001-11-13 & 1048 & 24.49 & 0.00 & 0.21 & 24.75 & 0.21 & 24.71 \\
2001-11-14 & 1049 & 24.48 & 0.00 & 0.22 & 24.74 & 0.22 & 24.72 \\
2001-11-15 & 1050 & 24.69 & $-$0.18 & 0.23 & 24.74 & 0.23 & 24.72 \\
2001-12-12 & 1077 & 24.80 & $-$0.25 & 0.22 & 24.76 & 0.22 & 24.74 \\
2002-02-01 & 1128 & 22.54 & $-$0.00 & 1.29 & 22.80 & 1.29 & 23.80 \\
2002-02-02 & 1129 & 24.72 & $-$0.25 & 0.26 & 24.67 & 0.27 & 24.69 \\
2002-02-03 & 1130 & 24.70 & $-$0.21 & 0.22 & 24.71 & 0.22 & 24.69 \\
2002-02-04 & 1131 & 24.63 & $-$0.18 & 0.21 & 24.68 & 0.21 & 24.65 \\
2002-02-05 & 1132 & 25.01 & $-$0.50 & 0.20 & 24.66 & 0.21 & 24.63 \\
2002-06-07 & 1254 & 24.85 & $-$0.30 & 0.19 & 24.75 & 0.19 & 24.69 \\
2002-10-12 & 1381 & 24.56 & $-$0.14 & 0.25 & 24.66 & 0.25 & 24.66 \\
2004-01-31 & 1857 & 24.41 & 0.00 & 0.22 & 24.67 & 0.22 & 24.64 \\
2004-02-01 & 1858 & 24.39 & 0.00 & 0.23 & 24.65 & 0.23 & 24.63 \\
%2004-03-19 & 1905 & 53.92 & $-$25.84 & 0.28 & 24.61 & 0.29 & 24.63 \\
2004-03-19 & 1905 & - & - & - & 24.61 & 0.29 & 24.63 \\
\hline
\end{tabular}
\end{table*}

\begin{table*}
  \caption{\label{tab:nights_V}
    Photometric solutions in the $V$-band. The colour term corresponds to the Johnson-Cousins colour $V-R$. The default value for the extinction coefficient in the one- and two-parameter fits is EXT1~$=$~EXT2~$=-0.11$ and the default value for the colour term in the one-parameter fit is CT1~$=-0.13$.}
\begin{tabular}{c r|r r r|r r|r }
\hline \hline
night & GaBoDS ID & ZP3 & EXT3 & CT3 & ZP2 & CT2 & ZP1 \\
\hline
%1999-11-08 & 312 & 30.51 & $-$5.48 & $-$0.16 & 24.37 & $-$0.16 & 24.35 \\
1999-11-08 & 312 & - & - & - & 24.37 & $-$0.16 & 24.35 \\
1999-12-02 & 336 & 24.49 & $-$0.33 & $-$0.12 & 24.27 & $-$0.12 & 24.27 \\
1999-12-03 & 337 & 24.34 & $-$0.17 & $-$0.18 & 24.27 & $-$0.18 & 24.24 \\
1999-12-04 & 338 & 24.23 & $-$0.07 & $-$0.17 & 24.27 & $-$0.17 & 24.25 \\
2000-03-29 & 454 & 24.24 & $-$0.13 & $-$0.17 & 24.22 & $-$0.17 & 24.20 \\
2000-03-30 & 455 & 24.40 & $-$0.30 & $-$0.12 & 24.17 & $-$0.11 & 24.18 \\
2000-03-31 & 456 & 24.51 & $-$0.42 & $-$0.12 & 24.13 & $-$0.11 & 24.14 \\
2000-04-01 & 457 & 23.89 & 0.00 & $-$0.07 & 24.02 & $-$0.07 & 24.05 \\
2000-04-06 & 462 & 24.09 & 0.00 & $-$0.13 & 24.22 & $-$0.14 & 24.22 \\
2000-07-03 & 550 & 22.81 & 0.00 & $-$0.93 & 22.94 & $-$0.93 & 22.12 \\
2000-11-28 & 698 & 24.09 & 0.00 & $-$0.16 & 24.23 & $-$0.16 & 24.21 \\
2000-11-29 & 699 & 24.30 & $-$0.17 & $-$0.16 & 24.22 & $-$0.16 & 24.20 \\
2001-02-02 & 764 & 24.08 & $-$0.10 & $-$0.15 & 24.10 & $-$0.16 & 24.09 \\
2001-02-20 & 782 & 24.39 & $-$0.34 & $-$0.15 & 24.08 & $-$0.13 & 24.08 \\
2001-02-23 & 785 & 24.07 & 0.00 & $-$0.16 & 24.21 & $-$0.16 & 24.19 \\
2001-02-25 & 787 & 24.25 & $-$0.19 & $-$0.16 & 24.14 & $-$0.16 & 24.13 \\
2001-02-26 & 788 & 24.15 & $-$0.09 & $-$0.15 & 24.17 & $-$0.15 & 24.16 \\
2001-03-26 & 816 & 24.53 & $-$0.45 & $-$0.18 & 23.99 & $-$0.04 & 24.04 \\
2001-03-27 & 817 & 24.04 & 0.00 & $-$0.17 & 24.17 & $-$0.17 & 24.15 \\
2001-06-25 & 907 & 24.12 & 0.00 & $-$0.15 & 24.26 & $-$0.16 & 24.24 \\
2001-06-27 & 909 & 24.42 & $-$0.22 & $-$0.18 & 24.25 & $-$0.15 & 24.23 \\
2001-06-29 & 911 & 24.28 & $-$0.11 & $-$0.16 & 24.28 & $-$0.16 & 24.25 \\
2001-07-26 & 938 & 24.36 & $-$0.13 & $-$0.19 & 24.34 & $-$0.19 & 24.27 \\
2001-08-21 & 964 & 24.31 & $-$0.13 & $-$0.14 & 24.28 & $-$0.14 & 24.27 \\
2001-08-22 & 965 & 24.30 & $-$0.15 & $-$0.13 & 24.25 & $-$0.14 & 24.24 \\
2001-08-23 & 966 & 25.26 & $-$0.94 & $-$0.17 & 24.25 & $-$0.14 & 24.24 \\
2001-11-12 & 1047 & 24.07 & $-$0.00 & $-$0.23 & 24.20 & $-$0.23 & 24.15 \\
2001-11-19 & 1054 & 24.24 & $-$0.05 & $-$0.20 & 24.31 & $-$0.20 & 24.28 \\
2001-12-08 & 1073 & 24.27 & $-$0.10 & $-$0.22 & 24.29 & $-$0.22 & 24.25 \\
2001-12-13 & 1078 & 24.35 & $-$0.16 & $-$0.16 & 24.28 & $-$0.15 & 24.26 \\
2002-02-09 & 1136 & 23.94 & 0.00 & $-$0.17 & 24.07 & $-$0.17 & 24.05 \\
2002-03-09 & 1164 & 24.08 & $-$0.09 & $-$0.14 & 24.10 & $-$0.14 & 24.09 \\
2002-10-06 & 1375 & 24.56 & $-$0.38 & $-$0.15 & 24.22 & $-$0.15 & 24.21 \\
2002-10-07 & 1376 & 24.34 & $-$0.21 & $-$0.13 & 24.22 & $-$0.13 & 24.22 \\
2002-10-09 & 1378 & 24.06 & 0.00 & $-$0.15 & 24.19 & $-$0.15 & 24.18 \\
2002-10-10 & 1379 & 24.30 & $-$0.18 & $-$0.16 & 24.20 & $-$0.15 & 24.19 \\
2002-10-12 & 1381 & 24.21 & $-$0.13 & $-$0.16 & 24.19 & $-$0.16 & 24.17 \\
2004-01-11 & 1837 & 24.18 & $-$0.12 & $-$0.15 & 24.17 & $-$0.15 & 24.15 \\
2004-01-15 & 1841 & 24.19 & $-$0.15 & $-$0.15 & 24.14 & $-$0.15 & 24.13 \\
2004-01-21 & 1847 & 24.15 & $-$0.11 & $-$0.15 & 24.15 & $-$0.15 & 24.14 \\
2004-10-02 & 2102 & 24.06 &    0.00 & $-$0.11 & 24.19 & $-$0.11 & 24.20 \\
2004-10-09 & 2107 & -     & -       & -       & 23.92 & $-$0.26 & 23.86 \\
2005-09-30 & 2465 & 24.24 & $-$0.10 & $-$0.17 & 24.25 & $-$0.17 & 24.21 \\
2005-10-30 & 2495 & 24.36 & $-$0.20 & $-$0.20 & 24.23 & $-$0.18 & 24.20 \\
\hline
\end{tabular}
\end{table*}

\begin{table*}
  \caption{\label{tab:nights_R}
    Photometric solutions in the $R$-band. The colour term corresponds to the Johnson-Cousins colour $V-R$. The default value for the extinction coefficient in the one- and two-parameter fits is EXT1~$=$~EXT2~$=-0.07$ and the default value for the colour term in the one-parameter fit is CT1~$=0.00$.}
\begin{tabular}{c r|r r r|r r|r }
\hline \hline
night & GaBoDS ID & ZP3 & EXT3 & CT3 & ZP2 & CT2 & ZP1 \\
\hline
1999-12-04 & 338 & 24.47 & 0.00 & $-$0.06 & 24.55 & $-$0.06 & 24.51 \\
2000-03-29 & 454 & 24.46 & 0.00 & $-$0.09 & 24.55 & $-$0.09 & 24.50 \\
2000-03-30 & 455 & 24.39 & 0.00 & $-$0.05 & 24.47 & $-$0.05 & 24.44 \\
2000-03-31 & 456 & 24.32 & 0.00 & $-$0.03 & 24.40 & $-$0.03 & 24.39 \\
2000-04-05 & 461 & 24.49 & $-$0.06 & $-$0.02 & 24.50 & $-$0.02 & 24.49 \\
2000-04-06 & 462 & 24.44 & 0.00 & $-$0.02 & 24.52 & $-$0.02 & 24.51 \\
2000-07-27 & 574 & 24.54 & 0.00 & $-$0.12 & 24.63 & $-$0.12 & 24.55 \\
2000-08-25 & 603 & 25.04 & $-$0.38 & $-$0.15 & 24.59 & $-$0.09 & 24.49 \\
2000-08-26 & 604 & 24.88 & $-$0.27 & $-$0.15 & 24.63 & $-$0.14 & 24.47 \\
2000-08-27 & 605 & 25.04 & $-$0.38 & $-$0.13 & 24.66 & $-$0.12 & 24.50 \\
2000-12-25 & 725 & 24.46 & 0.00 & $-$0.01 & 24.54 & $-$0.01 & 24.53 \\
2000-12-26 & 726 & 24.54 & $-$0.08 & $-$0.02 & 24.53 & $-$0.01 & 24.52 \\
2001-02-20 & 782 & 24.45 & $-$0.09 & $-$0.02 & 24.42 & $-$0.02 & 24.41 \\
2001-02-22 & 784 & 24.36 & 0.00 & $-$0.04 & 24.44 & $-$0.04 & 24.42 \\
2001-02-23 & 785 & 24.42 & 0.00 & $-$0.03 & 24.51 & $-$0.04 & 24.49 \\
2001-03-26 & 816 & 24.67 & $-$0.28 & $-$0.03 & 24.28 & 0.06 & 24.31 \\
2001-03-27 & 817 & 24.42 & 0.00 & $-$0.07 & 24.51 & $-$0.07 & 24.47 \\
2001-06-20 & 902 & 24.49 & 0.00 & $-$0.06 & 24.60 & $-$0.07 & 24.56 \\
2001-06-21 & 903 & 24.49 & 0.00 & $-$0.05 & 24.60 & $-$0.05 & 24.57 \\
2001-06-26 & 908 & 24.18 & 0.00 & 0.29 & 24.28 & 0.29 & 24.41 \\
2001-06-27 & 909 & 24.70 & $-$0.11 & $-$0.12 & 24.64 & $-$0.12 & 24.55 \\
2001-06-29 & 911 & 24.65 & $-$0.08 & $-$0.12 & 24.64 & $-$0.12 & 24.50 \\
2001-06-30 & 912 & 24.39 & 0.00 & $-$0.02 & 24.48 & $-$0.02 & 24.47 \\
2001-08-20 & 963 & 24.64 & $-$0.07 & $-$0.10 & 24.64 & $-$0.10 & 24.54 \\
2001-08-21 & 964 & 24.67 & $-$0.09 & $-$0.10 & 24.64 & $-$0.10 & 24.54 \\
2001-11-12 & 1047 & 24.39 & $-$0.00 & $-$0.01 & 24.47 & $-$0.01 & 24.47 \\
2001-11-16 & 1051 & 24.58 & $-$0.06 & $-$0.11 & 24.59 & $-$0.11 & 24.52 \\
2001-11-17 & 1052 & 24.62 & $-$0.11 & $-$0.09 & 24.58 & $-$0.10 & 24.53 \\
2001-11-19 & 1054 & 24.53 & 0.00 & $-$0.08 & 24.61 & $-$0.08 & 24.57 \\
2001-12-09 & 1074 & 24.50 & 0.00 & $-$0.09 & 24.58 & $-$0.09 & 24.54 \\
2001-12-11 & 1076 & 24.55 & $-$0.05 & $-$0.05 & 24.57 & $-$0.05 & 24.54 \\
%2003-04-05 & 1556 & 36.08 & $-$10.10 & $-$0.08 & 24.47 & $-$0.05 & 24.43 \\
2003-04-05 & 1556 & - & - & - & 24.47 & $-$0.05 & 24.43 \\
2003-04-06 & 1557 & 24.40 & 0.00 & $-$0.06 & 24.48 & $-$0.06 & 24.44 \\
2003-04-21 & 1572 & 24.58 & $-$0.08 & $-$0.12 & 24.56 & $-$0.12 & 24.51 \\
2004-10-04 & 2104 & 24.26 &    0.00 & $-$0.06 & 24.44 & $-$0.06 & 24.40 \\
2004-10-09 & 2109 & 24.32 &    0.00 & $-$0.08 & 24.40 & $-$0.08 & 24.36 \\
2004-10-10 & 2110 & -     & -       & -       & 24.56 & $-$0.12 & 24.49 \\
\hline
\end{tabular}
\end{table*}

\begin{table*}
  \caption{\label{tab:nights_I}
    Photometric solutions in the $I$-band. The colour term corresponds to the Johnson-Cousins colour $R-I$. The default value for the extinction coefficient in the one- and two-parameter fits is EXT1~$=$~EXT2~$=-0.10$ and the default value for the colour term in the one-parameter fit is CT1~$=0.11$.}
\begin{tabular}{c r|r r r|r r|r }
\hline \hline
night & GaBoDS ID & ZP3 & EXT3 & CT3 & ZP2 & CT2 & ZP1 \\
\hline
1999-11-04 & 308 & 23.16 & 0.00 & 0.28 & 23.27 & 0.28 & 23.37 \\
1999-11-07 & 311 & 26.18 & $-$2.61 & 0.27 & 23.31 & 0.27 & 23.40 \\
1999-11-08 & 312 & 23.24 & $-$0.00 & 0.24 & 23.36 & 0.24 & 23.44 \\
2000-02-26 & 422 & 22.97 & 0.00 & 0.29 & 23.09 & 0.28 & 23.18 \\
2000-03-29 & 454 & 25.13 & $-$1.66 & 0.16 & 23.11 & 0.10 & 23.11 \\
2000-03-30 & 455 & 23.59 & $-$0.50 & 0.28 & 23.12 & 0.27 & 23.21 \\
2000-03-31 & 456 & 22.87 & 0.00 & 0.32 & 22.99 & 0.32 & 23.11 \\
2000-04-01 & 457 & 22.99 & $-$0.00 & 0.22 & 23.11 & 0.22 & 23.17 \\
2000-04-05 & 461 & 23.00 & 0.00 & 0.28 & 23.13 & 0.28 & 23.22 \\
2000-04-06 & 462 & 23.19 & $-$0.12 & 0.31 & 23.17 & 0.31 & 23.28 \\
2000-07-04 & 551 & 23.10 & $-$0.04 & 0.35 & 23.16 & 0.35 & 23.44 \\
2000-07-28 & 575 & 23.04 & $-$0.00 & 0.23 & 23.16 & 0.23 & 23.24 \\
2000-07-31 & 578 & 23.06 & 0.00 & 0.26 & 23.18 & 0.26 & 23.28 \\
2000-08-01 & 579 & 23.08 & 0.00 & 0.28 & 23.21 & 0.28 & 23.32 \\
2000-08-03 & 581 & 22.81 & 0.00 & 0.26 & 22.92 & 0.26 & 23.02 \\
2000-11-29 & 699 & 23.13 & $-$0.09 & 0.31 & 23.15 & 0.31 & 23.26 \\
2000-12-25 & 725 & 23.04 & 0.00 & 0.32 & 23.15 & 0.32 & 23.27 \\
2000-12-26 & 726 & 23.04 & $-$0.01 & 0.31 & 23.17 & 0.30 & 23.28 \\
2001-02-21 & 783 & 23.07 & $-$0.18 & 0.28 & 22.97 & 0.29 & 23.06 \\
2001-03-24 & 814 & 23.09 & $-$0.11 & 0.29 & 23.08 & 0.28 & 23.18 \\
2001-03-25 & 815 & 22.95 & 0.00 & 0.29 & 23.08 & 0.29 & 23.18 \\
2001-06-19 & 901 & 23.15 & $-$0.09 & 0.29 & 23.17 & 0.29 & 23.28 \\
2001-06-20 & 902 & 23.77 & 0.00 & $-$1.05 & 23.88 & $-$0.99 & 23.29 \\
2001-06-21 & 903 & 23.18 & $-$0.09 & 0.29 & 23.20 & 0.29 & 23.30 \\
2001-06-22 & 904 & 23.29 & 0.00 & $-$1.02 & 23.40 & $-$1.01 & 22.77 \\
2001-06-24 & 906 & 23.02 & 0.00 & 0.34 & 23.14 & 0.36 & 23.32 \\
2001-06-25 & 907 & 23.10 & $-$0.08 & 0.37 & 23.13 & 0.37 & 23.31 \\
2001-06-26 & 908 & 23.04 & 0.00 & 0.27 & 23.17 & 0.29 & 23.27 \\
2001-06-29 & 911 & 23.21 & $-$0.14 & 0.38 & 23.16 & 0.37 & 23.46 \\
2001-06-30 & 912 & 23.05 & $-$0.08 & 0.43 & 23.08 & 0.43 & 23.39 \\
2001-07-20 & 932 & 25.32 & $-$1.97 & 0.48 & 22.75 & 0.57 & 23.31 \\
2001-07-21 & 933 & 22.59 & 0.00 & 0.60 & 22.73 & 0.59 & 23.27 \\
2001-07-22 & 934 & 23.08 & 0.00 & 0.33 & 23.20 & 0.34 & 23.45 \\
2001-07-23 & 935 & 23.64 & $-$0.50 & 0.35 & 23.11 & 0.35 & 23.41 \\
2001-07-24 & 936 & 23.07 & 0.00 & 0.34 & 23.19 & 0.34 & 23.44 \\
2001-07-25 & 937 & 23.07 & 0.00 & 0.36 & 23.21 & 0.36 & 23.52 \\
2001-07-26 & 938 & 23.13 & $-$0.05 & 0.35 & 23.20 & 0.35 & 23.50 \\
2001-07-27 & 939 & 22.66 & 0.00 & 0.40 & 22.81 & 0.40 & 23.18 \\
2001-08-19 & 962 & 23.08 & 0.00 & 0.25 & 23.21 & 0.25 & 23.29 \\
2001-08-21 & 964 & 23.00 & $-$0.02 & 0.41 & 23.10 & 0.42 & 23.41 \\
\hline
\end{tabular}
\end{table*}

\begin{table*}
  \caption{\label{tab:nights_I_EIS}
    Photometric solutions in the $I\_EIS$-band. The colour term corresponds to the Johnson-Cousins colour $R-I$. The default value for the extinction coefficient in the one- and two-parameter fits is EXT1~$=$~EXT2~$=0.00$ and the default value for the colour term in the one-parameter fit is CT1~$=0.03$.}
\begin{tabular}{c r|r r r|r r|r }
\hline \hline
night & GaBoDS ID & ZP3 & EXT3 & CT3 & ZP2 & CT2 & ZP1 \\
\hline
2002-03-10 & 1165 & 23.27 & 0.00 & 0.06 & 23.27 & 0.06 & 23.28 \\
2002-03-11 & 1166 & 23.33 & 0.00 & 0.06 & 23.33 & 0.06 & 23.35 \\
2002-03-12 & 1167 & 23.36 & 0.00 & 0.06 & 23.36 & 0.06 & 23.38 \\
2002-03-13 & 1168 & 23.39 & $-$0.02 & 0.05 & 23.37 & 0.06 & 23.38 \\
2005-07-28 & 2401 & 25.08 & $-$1.45 & 0.06 & 23.35 & 0.06 & 23.37 \\
\hline
\end{tabular}
\end{table*}

\section{Comparison to the EIS data}
In Table~\ref{tab:compare_EIS} the released data from the EIS team are
compared to our data.
\begin{table*}
  \caption{\label{tab:compare_EIS}
    Comparison of the GaBoDS (coaddition ID ``D??A'' and for the Deep2c $U$- and $V$-band images ``D2CB'', respectively) and the EIS data. The mean and the standard deviation of the magnitude differences ($\Delta m=m_{\mathrm{GaBoDS}}-m_{\mathrm{EIS}}$) are computed for objects in the magnitude range $17<m<21$. Due to the mistakes in the EIS photometric calibration (see text), in column four only the numbers printed in bold face are meaningful.}
\begin{tabular}{l l l r r r }
\hline \hline
Field & Filter & EIS ID & $\Delta m$ [mag] & $\cos(\delta)\cdot\Delta \alpha$ [$\arcsec$] & $\Delta\delta$ [$\arcsec$]\\
\hline
Deep1a & $U\_35060$ & \verb_EIS.2004-11-05T11:29:36.834_ & $-0.188\pm\mathbf{0.040}$ & $0.045\pm0.117$ & $0.034\pm0.081$ \\
Deep1a & $U\_38$ & \verb_EIS.2004-11-05T13:29:55.700_ & $-0.172\pm\mathbf{0.043}$ & $0.035\pm0.121$ & $0.043\pm0.079$ \\
Deep1a & $B$ & \verb_EIS.2004-11-04T20:33:40.955_ & $-0.235\pm\mathbf{0.048}$ & $-0.001\pm0.111$ & $0.042\pm0.079$ \\
Deep1a & $R$ & \verb_EIS.2004-11-04T20:41:28.432_ & $-0.016\pm\mathbf{0.036}$ & $0.002\pm0.119$ & $0.015\pm0.079$ \\
Deep1a & $I$ & \verb_EIS.2004-11-04T21:20:50.202_ & $-0.186\pm\mathbf{0.051}$ & $0.001\pm0.118$ & $0.021\pm0.075$ \\
Deep1b & $U\_35060$ & \verb_EIS.2004-11-05T13:52:48.142_ & $-0.330\pm\mathbf{0.109}$ & $0.084\pm0.110$ & $-0.005\pm0.085$ \\
Deep1b & $B$ & \verb_EIS.2004-11-04T20:56:47.504_ & $-0.320\pm\mathbf{0.068}$ & $0.068\pm0.098$ & $-0.016\pm0.069$ \\
Deep1b & $V$ & \verb_EIS.2004-11-04T20:33:41.505_ & $-0.030\pm\mathbf{0.034}$ & $0.035\pm0.091$ & $-0.028\pm0.064$ \\
Deep1b & $R$ & \verb_EIS.2004-11-04T21:02:53.765_ & $-0.124\pm\mathbf{0.038}$ & $0.041\pm0.096$ & $-0.017\pm0.060$ \\
Deep1b & $I$ & \verb_EIS.2004-11-04T21:55:23.563_ & $-0.137\pm\mathbf{0.048}$ & $0.047\pm0.087$ & $-0.018\pm0.058$ \\
Deep1c & $V$ & \verb_EIS.2004-11-04T20:58:46.283_ & $-0.027\pm\mathbf{0.085}$ & $-0.015\pm0.153$ & $-0.035\pm0.070$ \\
Deep1c & $R$ & \verb_EIS.2004-11-04T21:45:25.126_ & $-0.108\pm\mathbf{0.032}$ & $-0.043\pm0.146$ & $-0.043\pm0.071$ \\
Deep2a & $R$ & \verb_EIS.2004-11-04T22:09:46.854_ & $-0.094\pm0.082$ & $-0.153\pm0.121$ & $-0.088\pm0.132$ \\
Deep2b & $U\_35060$ & \verb_EIS.2004-11-05T14:52:18.150_ & $-0.015\pm0.106$ & $-0.119\pm0.129$ & $0.090\pm0.107$ \\
Deep2b & $B$ & \verb_EIS.2004-11-04T21:29:26.100_ & $-0.309\pm0.094$ & $-0.183\pm0.110$ & $0.041\pm0.091$ \\
Deep2b & $V$ & \verb_EIS.2004-11-04T21:15:08.835_ & $-0.034\pm0.092$ & $-0.187\pm0.111$ & $0.030\pm0.088$ \\
Deep2b & $R$ & \verb_EIS.2004-11-04T22:25:36.371_ & $-0.104\pm0.089$ & $-0.196\pm0.121$ & $0.043\pm0.095$ \\
Deep2b & $I$ & \verb_EIS.2004-11-04T22:49:00.123_ & $-0.207\pm0.116$ & $-0.203\pm0.110$ & $0.025\pm0.098$ \\
Deep2c & $U\_35060$ & \verb_EIS.2004-11-05T16:06:26.115_ & $-0.486\pm0.124$ & $-0.118\pm0.116$ & $0.067\pm0.096$ \\
Deep2c & $V$ & \verb_EIS.2004-11-04T21:34:51.734_ & $-0.286\pm0.134$ & $-0.168\pm0.126$ & $0.018\pm0.122$ \\
Deep2c & $I$ & \verb_EIS.2004-11-04T23:17:21.542_ & $-0.132\pm0.078$ & $-0.163\pm0.106$ & $0.007\pm0.094$ \\
Deep2d & $R$ & \verb_EIS.2004-11-04T22:40:01.850_ & $-0.315\pm0.080$ & $-0.189\pm0.105$ & $0.028\pm0.075$ \\
Deep3a & $U\_35060$ & \verb_EIS.2004-11-05T18:10:27.273_ & $-0.200\pm\mathbf{0.039}$ & $-0.209\pm0.147$ & $-0.393\pm0.144$ \\
Deep3a & $U\_38$ & \verb_EIS.2004-11-05T18:54:07.084_ & $-0.016\pm\mathbf{0.044}$ & $-0.208\pm0.162$ & $-0.419\pm0.141$ \\
Deep3a & $B$ & \verb_EIS.2004-11-04T21:55:20.499_ & $-0.196\pm\mathbf{0.055}$ & $-0.233\pm0.125$ & $-0.417\pm0.130$ \\
Deep3a & $V$ & \verb_EIS.2004-11-04T22:03:21.246_ & $0.050\pm\mathbf{0.050}$ & $-0.219\pm0.126$ & $-0.434\pm0.129$ \\
Deep3a & $R$ & \verb_EIS.2004-11-04T22:49:41.273_ & $-0.084\pm\mathbf{0.086}$ & $-0.236\pm0.114$ & $-0.431\pm0.134$ \\
Deep3b & $U\_35060$ & \verb_EIS.2004-11-05T19:25:00.562_ & $0.093\pm\mathbf{0.071}$ & $-0.239\pm0.074$ & $-0.306\pm0.079$ \\
Deep3b & $B$ & \verb_EIS.2004-11-04T22:52:21.098_ & $-0.074\pm\mathbf{0.029}$ & $-0.234\pm0.074$ & $-0.326\pm0.060$ \\
Deep3b & $V$ & \verb_EIS.2004-11-04T22:31:19.567_ & $0.012\pm\mathbf{0.031}$ & $-0.234\pm0.071$ & $-0.329\pm0.060$ \\
Deep3b & $R$ & \verb_EIS.2004-11-04T23:18:23.026_ & $-0.076\pm\mathbf{0.033}$ & $-0.226\pm0.076$ & $-0.335\pm0.061$ \\
Deep3b & $I$ & \verb_EIS.2004-11-05T00:03:54.812_ & $-0.186\pm\mathbf{0.046}$ & $-0.239\pm0.069$ & $-0.337\pm0.054$ \\
Deep3c & $U\_35060$ & \verb_EIS.2004-11-05T20:05:53.946_ & $-0.323\pm\mathbf{0.076}$ & $-0.302\pm0.085$ & $-0.319\pm0.071$ \\
Deep3c & $B$ & \verb_EIS.2004-11-04T23:24:11.875_ & $-0.191\pm\mathbf{0.042}$ & $-0.317\pm0.083$ & $-0.332\pm0.058$ \\
Deep3c & $V$ & \verb_EIS.2004-11-04T22:52:44.317_ & $-0.022\pm\mathbf{0.029}$ & $-0.323\pm0.082$ & $-0.335\pm0.061$ \\
Deep3c & $R$ & \verb_EIS.2004-11-04T23:29:13.696_ & $-0.060\pm\mathbf{0.029}$ & $-0.312\pm0.084$ & $-0.341\pm0.059$ \\
Deep3c & $I$ & \verb_EIS.2004-11-05T00:33:52.446_ & $-0.081\pm\mathbf{0.044}$ & $-0.334\pm0.082$ & $-0.350\pm0.060$ \\
Deep3d & $V$ & \verb_EIS.2004-11-04T23:09:20.349_ & $-0.081\pm\mathbf{0.065}$ & $-0.367\pm0.070$ & $-0.342\pm0.065$ \\
Deep3d & $I\_EIS$ & \verb_EIS.2004-11-05T01:14:19.620_ & $-0.104\pm\mathbf{0.035}$ & $-0.363\pm0.064$ & $-0.353\pm0.056$ \\
\hline
\end{tabular}
\end{table*}

\section{FITS header}
\label{sec:header}
\begin{table*}
\caption{\label{tab:header}FITS header keywords for the released images. The astrometric keywords from RADECSYS to CDELT2 are described in \cite{2002A&A...395.1061G}.}
\begin{tabular}{l l l}
\hline \hline
Keyword  & Unit & comment\\
\hline
RADECSY  &      & astrometric system\\
CTYPE1   &      & WCS projection type for x-axis\\
CUNIT1   &      & x-axis unit\\
CRVAL1   &  deg & world x-coordinate of reference pixel\\
CRPIX1   &  pix & reference pixel on x-axis\\
CDELT1   &  deg/pix & pixel step along x-axis\\
CTYPE2   &      & WCS projection type for y-axis\\
CUNIT2   &      & y-axis unit\\
CRVAL2   &  deg & world y-coordinate of reference pixel\\
CRPIX2   &  pix & reference pixel on y-axis\\
CDELT2   &  deg/pix & pixel step along y-axis\\
EXPTIME  &  sec & total exposure time\\
GAIN     &      & effective GAIN (instrumental gain [2.0] $\times$ EXPTIME)\\
MAGZP    &  mag & Vega magnitude zeropoint\\
SEEING   &  arcsec & measured image seeing\\
COND1-5  &      & condition on the input images entering coaddition\\
NIGHT1-3 &      & GaBoDS IDs and solutions for nights included in phot. solution\\
\hline
\end{tabular}
\end{table*}

In the following we show an excerpt of one of our image headers to
describe the special keywords inserted by the reduction
pipeline\footnote{The image headers of the files released via the ESO
  archive at
  \url{http://archive.eso.org/archive/eso_data_products.html} are
  slightly different and contain a lot of additional keywords in order
  to comply with archive standards.}. EXPTIME is the sum of the
exposure times of all images that entered the coaddition. GAIN is the
instrumental gain (2.0) multiplied by EXPTIME. MAGZP is the Vega
magnitude zeropoint which is to be used for converting counts into
magnitudes by:
$\mathrm{mag}=\mathrm{MAGZP}-2.5\cdot\log\left(\mathrm{counts}\right)$.
The COND keywords contain the filter conditions that were applied to
the image catalogues before coaddition. Within these conditions
AUTO\_ZP represents a filtering on the single images' relative
zeropoints, BACKGR a filtering on the single images' background flux,
and SEEING a filtering on the single images' measured seeing. The
NIGHT keywords summarise the absolute photometric calibration
containing pairs of GaBoDS IDs and chosen solutions, e.g. 903 2 means
that in the night number 903 (2001-06-21) the two-parameter fit was
chosen.

In Table~\ref{tab:header} the important FITS header keywords of our
released images are summarised.

\begin{verbatim}
====> file Deep3c/Deep3c_I.D3CA.swarp.fits (main) <====
...
...
...
EXPTIME =                25193 / total Exposure Time
GAIN    =                50386 / effective GAIN for SExtractor
MAGZP   =              23.0005 / Vega Magnitude Zeropoint
SEEING  =              1.02218 / measured image Seeing
COMMENT
COMMENT Conditions on the input images:
COND1   = '((SEEING<2)AND(AUTO_'
COND2   = 'ZP>0));'
DUMMY8  =                    0 / DUMMY keyword
DUMMY9  =                    0 / DUMMY keyword
DUMMY10 =                    0 / DUMMY keyword
NIGHT1  = '903 2, 906 2, 907 2,'
NIGHT2  = '908 2, 911 2,'
NIGHT3  = ', 911 0, 912 0,'
...
...
...
\end{verbatim}

\end{document}